\pdfoutput=1

\documentclass[a4paper,11pt]{article}

\usepackage{jheppub,fancyhdr,epstopdf,color,amsmath,cases,slashed,hyperref,subfigure}
\usepackage{amsfonts}
\usepackage{amssymb}
\usepackage{graphicx,graphics,epsfig}
\usepackage{geometry}
\usepackage[english]{babel}
\usepackage{mathtools}
\usepackage[utf8]{inputenc}
\usepackage[T1]{fontenc}
 \usepackage{float}
\usepackage{hyperref}
\usepackage{caption}
\usepackage{caption}
\usepackage{subfigure}  
\usepackage{mathrsfs}  
\usepackage{bbold}
\usepackage{comment}
\usepackage{float}
\usepackage{stackrel}
\usepackage{multirow}
\usepackage{slashed}

\def\noi{\noindent}

\def\l{\lambda}

\def\ba{\begin{array}}
\def\ea{\end{array}}
\def\bea{\begin{eqnarray}}
\def\eea{\end{eqnarray}}

\newcommand{\Asla}{\not{\hbox{\kern-3.5pt $A$}}}
\newcommand{\Gsla}{\not{\hbox{\kern-3.5pt $G$}}}
\newcommand{\Wsla}{\not{\hbox{\kern-3.5pt $W$}}}
\newcommand{\Zsla}{\not{\hbox{\kern-3.5pt $Z$}}}

\newcommand{\Dslash}{\not{\hbox{\kern-4pt $D$}}}
\newcommand{\pslash}{\not{\hbox{\kern-2.3pt $p$}}}

\newcommand{\pihhm}{\Pi_{hh}(M_h^2)}
\newcommand{\pihhs}{\Pi_{hh}(s)}

\newcommand{\cpihhs}{\hat{\Pi}_{hh}(s)}

\newcommand{\ra}{\rightarrow}
\def\lsim{\;\raise0.3ex\hbox{$<$\kern-0.75em\raise-1.1ex\hbox{$\sim$}}\;}
\def\gsim{\;\raise0.3ex\hbox{$>$\kern-0.75em\raise-1.1ex\hbox{$\sim$}}\;}

\def\l{\lambda}

\def\ba{\begin{array}}
\def\ea{\end{array}}
\def\bea{\begin{eqnarray}}
\def\eea{\end{eqnarray}}

\def\bll{\tilde{\beta}_{\l_L}}

\def\lsim{\;\raise0.3ex\hbox{$<$\kern-0.75em\raise-1.1ex\hbox{$\sim$}}\;}
\def\gsim{\;\raise0.3ex\hbox{$>$\kern-0.75em\raise-1.1ex\hbox{$\sim$}}\;}

\newcommand{\beqn}{\begin{eqnarray}}
\newcommand{\eeqn}{\end{eqnarray}}

\title{Relic density of dark matter in the inert doublet model beyond leading order for the low mass region:  4. The  Higgs resonance region}

\preprint{LAPTH-004/21, CERN-TH-2021-004}

\author[a]{Shankha Banerjee}
\author[b]{\!\!, Fawzi Boudjema}
\author[c, d]{\!\!, Nabarun Chakrabarty}
\author[e]{\!\!,  Hao Sun}

\affiliation[a]{CERN, Theoretical Physics Department, CH-1211 Geneva 23, Switzerland}
\affiliation[b]{LAPTh, Universit\'e Savoie Mont Blanc, CNRS, BP~110, F-74941 Annecy-le-Vieux, France}
\affiliation[c]{Centre for High Energy Physics, Indian Institute of Science, C.V. Raman Avenue, Bangalore 560012, India}
\affiliation[d]{Department of Physics, Indian Institute of Technology Kanpur, Kanpur, Uttar Pradesh 208016, India}
\affiliation[e]{Institute of Theoretical Physics, School of Physics, Dalian University of Technology, Dalian 116024, People’s Republic of China}

\emailAdd{shankha.banerjee@cern.ch}
\emailAdd{boudjema@lapth.cnrs.fr}
\emailAdd{chakrabartynabarun@gmail.com}
\emailAdd{haosun@dlut.edu.cn}

\abstract{One-loop electroweak corrections to the annihilation cross-sections of dark matter in the Higgs resonance region of the inert doublet model (IDM) are investigated. The procedure of how to implement the width of the Higgs in order to regularise the amplitude both at tree-level and at one-loop together with the renormalisation of a key parameter of the model, are thoroughly scrutinised. The discussions go beyond the application to the relic density calculation and also beyond the IDM so that addressing these technical issues can help in a wider context. We look in particular at the dominant channels with the $b \bar b$ final state and the more involved 3-body final state, $W f \bar f^\prime$, where both a resonance and an anti-resonance, due to interference effects, are present. We also discuss how to integrate over such configurations when converting the cross-sections into a calculation of the relic density.}

\begin{document}

\date\today

\maketitle


\section{Introduction}
\label{sec:introHiggsRes}
There is a fine-tuned region in the parameter space of the inert doublet model (IDM)~\cite{Deshpande:1977rw, Barbieri:2006dq,Hambye:2007vf, LopezHonorez:2006gr,Cao:2007rm,Gustafsson:2007pc,Agrawal:2008xz,Hambye:2009pw, Lundstrom:2008ai, Andreas:2009hj, Arina:2009um, Dolle:2009ft, Nezri:2009jd, Miao:2010rg, Gong:2012ri, Gustafsson:2012aj, Swiezewska:2012eh, Arhrib:2012ia,Wang:2012zv, Goudelis:2013uca, Arhrib:2013ela, Krawczyk:2013jta, Osland:2013sla, Abe:2015rja,Blinov:2015qva, Diaz:2015pyv, Ilnicka:2015jba, Belanger:2015kga, Carmona:2015haa, Kanemura:2016sos,Queiroz:2015utg,Belyaev:2016lok,Arcadi:2019lka, Eiteneuer:2017hoh, Ilnicka:2018def, Kalinowski:2018ylg,Ferreira:2009jb,Ferreira:2015pfi,Kanemura:2002vm, Senaha:2018xek, Braathen:2019pxr, Arhrib:2015hoa,Garcia-Cely:2015khw,Banerjee:2016vrp,Basu:2020qoe,Abouabid:2020eik,Kalinowski:2020rmb} where if the mass of the lightest neutral scalar of the model $X$ is such that $2 M_X \simeq M_h=125$GeV ($h$ is the Standard Model (SM) Higgs), a good value of the relic density is obtained. This is possible, thanks to the efficient annihilation of the DM pair through, essentially, the SM Higgs resonance. The latter couples to the DM pair with a coupling of strength, $\l_L$. In Ref.~\cite{OurPaper1_2020} where an extensive analysis of the parameter space of the model is conducted, taking into account various experimental and theoretical constraints, two benchmarks points for the Higgs resonance region are selected. We list the characteristics of these 2 points in Table~\ref{tab:p57p59}.

\begin{table}[hbtp]
\centering
\begin{tabular}{|c|c |c |}
\cline{2-3}
\multicolumn{1}{c|}{}& P57&P59  \\
\hline
$M_X$  &   57&59\\ 
\hline 
$\l_L \times 10^3$ &2.4 & 1.0  \\ \hline
$M_A$,$M_{H^\pm}$&      113,123&113,123 \\  \hline 
$( \l_3,\l_4,\l_5)$ & (0.382, -0.228,-0.152) & (0.373,-0.224,-0.148)\\ 
\hline$\bll$ & $1.123+ 2.157 \; \l_2      $& $1.097 + 2.095 \; \l_2 $  \\ \hline
\hline
\multicolumn{3}{|c|}{$\Omega h^2$} \\    \hline
$\alpha(0)$ &  0.113&0.108  \\    \hline
$\alpha(M_Z^2)$ &  0.118&0.113  \\    \hline
$\Omega_{b \bar b}(\%) $     &58&57 \\ \hline
$\Omega_{WW^\star}(\%)$ &22&24 \\ \hline     
$\Omega_{gg}(\%)$ &8 & 7 \\ \hline
\hline
$\alpha(M_Z^2),\l_L=0$ &  1.97&1.97  \\    \hline     
\hline
\end{tabular}
\caption{\label{tab:p57p59}{\it Characteristics of the benchmark points for the Higgs resonance region, P57 and P59. $H^\pm$ are the charged scalars of the IDM. Beside the DM scalar, $X$, $A$ is the second neutral scalar often referred to as the pseudo-scalar of the IDM. All masses are in GeV. $\l_{2,3,4,5}$ are the parameters of the scalar potential of the IDM. The tree-level relic density (calculated with $\alpha(0) $, the electromagnetic coupling in the Thomson limit)  and the weight in percent for all channels contributing to more than $5\%$ of the relic density, are shown. These are  the dominant $b\bar b$ bottom quark pair final state, $WW^\star$ and gluon pair, $gg$. The last line gives the value of the relic density when $\l_L=0$, the Higgs resonance being thus switched off. We also list the values of the underlying parameters $\l_{3,4,5}$ and $\bll$ (the one-loop $\beta$ constant associated to $\l_L$)~\cite{OurPaper1_2020}. We use {\tt micrOMEGAs 5.0.7}~\cite{Belanger:2001fz, Belanger:2004yn, Belanger:2006is, Belanger:2013oya, Belanger:2018mqt} to compute the relic density.}}
\end{table}

The prominence of the Higgs resonance is made obvious upon switching off Higgs exchange by setting $\l_L=0$. With $\l_L=0$, the relic density is almost as much as $20$ times larger. This huge increase is due to the extremely small residual gauge contribution from $XX \to WW^\star (Z Z^\star)$ through the $t$-channel exchange of $H^\pm(A)$, which we study at length in~\cite{OurPaper1_2020}. Barring this residual contribution, the percentages of the participating cross-sections to the relic density in Table~\ref{tab:p57p59} correspond to, a very good approximation~\footnote{With the tiny values of $\l_L$ for both P57 and P59, ${\rm Br}(h\to XX)< 0.3\%$ (see Ref.~\cite{OurPaper1_2020}). This very small branching into invisibles does not affect much the SM values of the Higgs partial widths.} 
the branching fraction of the SM Higgs to the corresponding channels including the $gg$ (gluon pair final state). Only the $W W^\star= W f \bar f^\prime$ channel for the relic density is $2\%$ larger than the partial decay fraction of the Higgs to $W W^\star$ due to the small continuum gauge contribution, $XX\to WW^\star$, for the relic. The total width and the branching fractions are therefore key parameters for describing the Higgs lineshape and the computation of the relic density in the Higgs resonance region. Small changes in or very close to the vicinity of the resonance could lead to large corrections to the cross-sections and hence to the relic density. If one observes that the non resonant $WW$ channel contributes even in the absence of Higgs exchange, one could ask, for instance, if other contributions to the same channels could be induced at one-loop, away from the resonance, impacting the cross-sections. Because of the inert nature of the model, annihilation to fermions only takes place through the Higgs exchange at tree-level. However, at electroweak one-loop corrections these annihilations proceed even when the Higgs exchange is absent. Nonetheless, because of the small width of the SM Higgs, the properties of the Higgs resonance are crucial. Apart from the mass of the Higgs, they rely on the total width and the branching fractions of the SM Higgs. \\
The plan of the paper is as follows. We first review the key parameters, in particular the SM parameters that enter the description of the Higgs resonance. Since the paper deals with the electroweak corrections, the discussion about the parameters, in particular the bottom-quark mass, is covered in the next section which aims to describe how the important QCD corrections are included. Since, as known and as we will argue, the width is a higher order effect, its introduction at tree-level and at one-loop should be treated with care, in particular in order to avoid double counting of corrections and also, in general, and how to avoid breaking gauge invariance. In this spirit, section~\ref{sec:treebb} presents the tree-level calculation of the $b\bar b$ final state through the Higgs resonance. The Breit-Wigner parametrisation will prove to be most useful when we deal with the one-loop electroweak correction which will be covered in Section~\ref{sec:oneloopbb}. We will see how the perturbation series needs to be reorganised in order to avoid the issue of double counting. The next 2 sections, Section~\ref{sec:treehWW} and Section~\ref{sec:loophWW}, will follow similar steps in the case of $W W^\star=W f \bar f^\prime$. Some important technical issues need to be tackled here. The $W W^\star=W f \bar f^\prime$ final state features a resonance contribution and also an interference with a "continuum" part. The whole effect will exhibit a resonance and "anti-resonance" behaviour over a small range of energy. Section~\ref{sec:applicationtorelic} turns the corrected cross-sections into the improved relic density. We show how we must be extra careful in integrating the cross-sections over the velocity distribution needed in this conversion. Our conclusions are presented in Section~\ref{sec:conclusions}.


\section{The inputs and the parameters to study the resonance region}
\label{sec:parameters}
All the partial widths and total widths can be derived from theory. The latest state-of-the art calculations~\cite{deFlorian:2016spz, 10.1093/ptep/ptaa104} for the total SM width of the Higgs gives the value 
\beqn
\Gamma_h=4.07 \; {\rm MeV} \;\; ({\pm 4\%}).
\eeqn 
Considering the theoretical error on the total width (much larger than that contribution due to $\Gamma(h\to XX)$ after imposing the direct detection constraint on $\l_L$\cite{OurPaper1_2020}), we equate the total width with the SM total width. These most precise calculations of the SM Higgs width include crucial higher order QCD corrections. They also lead to the theoretical prediction for the branching ratio into $b\bar b$ to be $58.2\%$, which corresponds to $\Gamma(h \to b \bar b)=2.37$ MeV. In order to incorporate these important QCD higher order predictions in an electroweak calculation at tree-level we will use, in this paper, an effective $m_{b,\rm eff.}$ for the Higgs coupling such that 
the correct value of the branching width into $b\bar b$ is recovered. With the tree-level prediction of the $b\bar b$ partial width given by 
\beqn
\label{eq:Gammahtobb}
\Gamma(h \to b \bar b)=M_h \frac{N_c}{8\pi} \frac{m_{b,\rm eff.}^2}{v^2} \beta_f, \quad \beta_f=\sqrt{1-4 m_{b,\rm eff.}^2/M_h^2}, \quad N_c=3, \quad M_W=\frac{g v}{2}.
\eeqn
$M_W$ is the $W$-boson mass ($g$ is the $SU(2)$ coupling and $v$ the vacuum expectation value). We therefore take for the Higgs resonance channel $m_b=3.15375$ GeV as the {\it effective} mass to incorporate the important QCD corrections~\footnote{Note that this effective mass is not to be used as the {\em input} mass in {\tt micrOMEGAs}. For {\tt micrOMEGAs}, a default $b$-pole mass is used as input. Internally, the code calls different routines for running masses and QCD corrections to the $hbb$ vertex.}. In our calculation of the cross-sections for these benchmark points, we therefore only consider the  electroweak corrections, that unlike the QCD correction, affect both the initial and final states. We should also keep in mind that $\Gamma(h \to b \bar b)$ in Equation~\ref{eq:Gammahtobb} does not include the electroweak radiative corrections which as we will see are quite small. 

Apart from the numerical value of the Higgs width, we have more to say about how we implement the Higgs width in our calculations. It is sufficient to say in this preamble that we take the total SM Higgs width as an input (at all orders), the same way as we treat the Higgs mass as an input (at all orders) in perturbation. As Table~\ref{tab:p57p59} shows, the most important channels are by far the $b \bar b$ and the $WW$. Although the sum of the rest of the individual channels is not at all negligible ($20\%$), all fermion pair final states  proceed through the Higgs exchange similarly to the $b\bar b$ channel and hence the procedure of applying the electroweak radiative corrections to them is the same as in the $b \bar b$ channel. The contribution of the $ZZ$ final state is below $3\%$, which is too small a contribution for any correction to make an impact on the relic density calculation, especially in view of the results we find in the $WW$ channel. Technically, $ZZ$ involves the same treatment as the annihilation to $WW$. Lastly, at the order we are performing the corrections, the Higgs exchange induced annihilation into gluon pairs will not be subject to electroweak corrections. 


\section{$X X \stackrel{h}{\longrightarrow} b \bar b$ at tree-level}
\label{sec:treebb}
\begin{figure}[htbp]
\begin{center}
\includegraphics[scale=0.35]{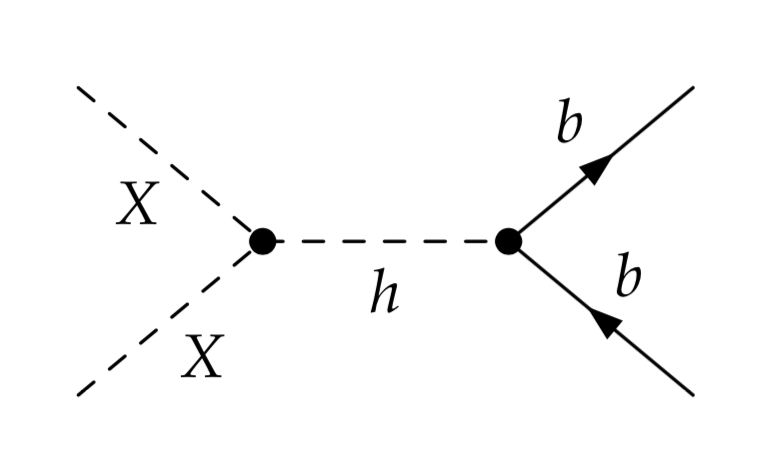}
\caption{\it  A single diagram, Higgs mediated, contributes to $XX \to b \bar b$ at tree-level.}
\label{fig:FeynXXbbtree}
\end{center}
\end{figure}
\begin{figure}[htbp]
\begin{center}
\includegraphics[scale=0.65]{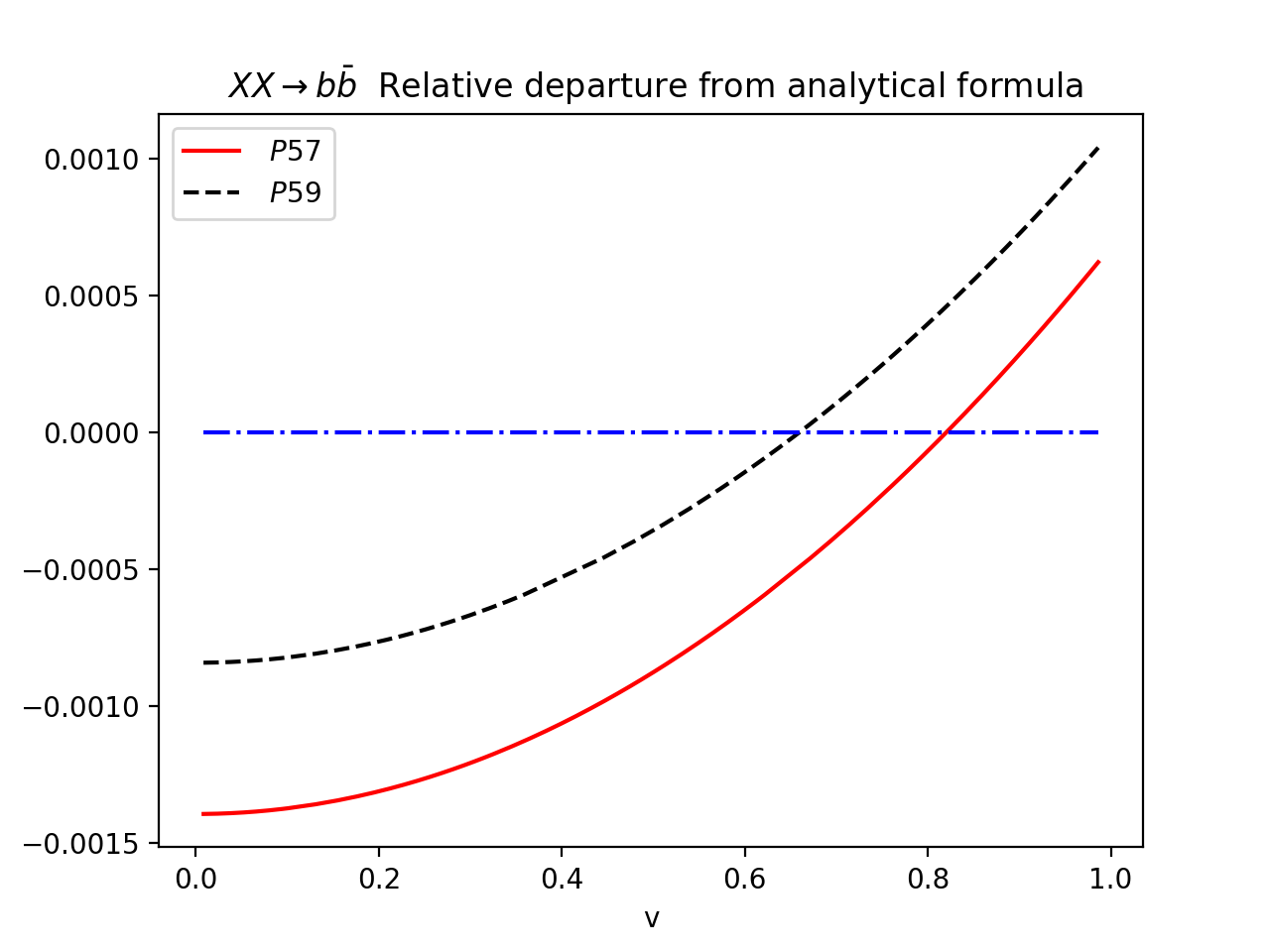}
\caption{\it  \label{fig:xxtobbtreeratio} The velocity dependence of the ratio of the full tree-level computation relative to the fixed width Breit-Wigner calculation based on Equation~\ref{eq:treeXXbb} for Points P57 and P59. The horizontal line, exact agreement, meets the curves at exactly the resonance points.}
\end{center}
\end{figure}

 
In the IDM, the tree-level annihilation to $b\bar b$ only occurs through the SM Higgs. In this case, it is relatively easy to parameterise the {\it tree-level} cross-section $XX \to b \bar b$ in a compact form as
\beqn
\label{eq:treeXXbb}
\sigma^0{(X X \stackrel{h}{\longrightarrow} b \bar b})\; v=
\frac{ 64 \; \pi \frac{\Gamma(h\ra X X)}{\sqrt{1-4M_X^2/M_h^2}} \Gamma(h\ra b\bar{b})}{(s -M_h^2)^2+
\Gamma_h^2 M_h^2}.
\eeqn
Here and in the following $^0$ refers to the tree-level cross-section. $s$, the invariant mass of the system, can be expressed in terms of the relative velocity $v$, $\sqrt{s}=2 M_X /\sqrt{1-\frac{v^2}{4}}$. Note that this tree-level cross-section is written solely in terms of physical observables, the total width and the partial widths into $XX$ and $b \bar b$. Observe also that in this Breit-Wigner cross-section, the total width and the partial widths are physical (on-shell) widths that do not run with $s$. It is therefore a totally gauge invariant cross-section. We take the partial width $\Gamma(h \to XX)$ as an {\it input} parameter in lieu of $\l_L$. We use the total width of the Higgs, $\Gamma_h$, as an input, making sure that the induced width that is generated at one-loop (the imaginary part of the Higgs two-point self-energy function) is properly subtracted (see later). For $\Gamma(h\ra b\bar{b})$, we stress again that we use the effective $b$-quark mass which is consistent with the value of $\Gamma(h \ra b\bar{b})$ (Equation~\ref{eq:Gammahtobb}), therefore both $\Gamma_h$ and $\Gamma(h\ra b\bar{b})$ include important resummed QCD corrections which we do not have to calculate when we calculate the electroweak corrections. The latter will also include, at one-loop, non resonant Higgs contributions. For $\Gamma(h\ra X X)$, we use Ref.~\cite{OurPaper1_2020}
\beqn
\label{eq:pwhinv}
\Gamma_{h\to XX}&=&\frac{\lambda_L^2 v^2}{32 \pi M_h} \sqrt{1-\frac{4 M_X^2}{M_h^2}}.
\eeqn
It is very important to observe that in the product $\Gamma(h \ra X X) \Gamma(h\ra b\bar{b})$, the value of the vacuum expectation value, $v$, drops out. Therefore, the dependence on the electromagnetic coupling that separately enters these partial widths through $v$ cancels out. As a result, with $\Gamma_h$ as an input parameter, the cross-section, $XX \to b \bar b$, does not depend on the value of the electromagnetic coupling $\alpha$. This is the reason the values of the relic density in Table~\ref{tab:p57p59} for the resonance region show quite negligible difference between the use of $\alpha(0)$ and $\alpha(M_Z^2)$ (the small difference comes from the small gauge contribution to the {\it continuum} $W W^\star$). \\
In our analysis, the cross-sections are exactly computed with the help of our automated code {\tt SloopS}~\cite{Baro:2007em, Baro:2008bg, Baro:2009na}. We verify that for the tree-level annihilation into $b \bar b$, the parameterisation of Equation~\ref{eq:treeXXbb} reproduces the results of the full calculation over the whole $v$ range with an agreement better than $5\times 10^{-4}$ around the resonance region, which is the region that matters most for the relic density calculation, as shown in Figure~\ref{fig:xxtobbtreeratio}.

This verification was in order not only to understand the features of the resonance processes but also to seek effective implementation at one-loop, later. We have more to say about equation~\ref{eq:treeXXbb} after discussing the set up of the loop calculation to this observable.

\begin{figure}[htbp]
\begin{center}
\includegraphics[scale=0.50]{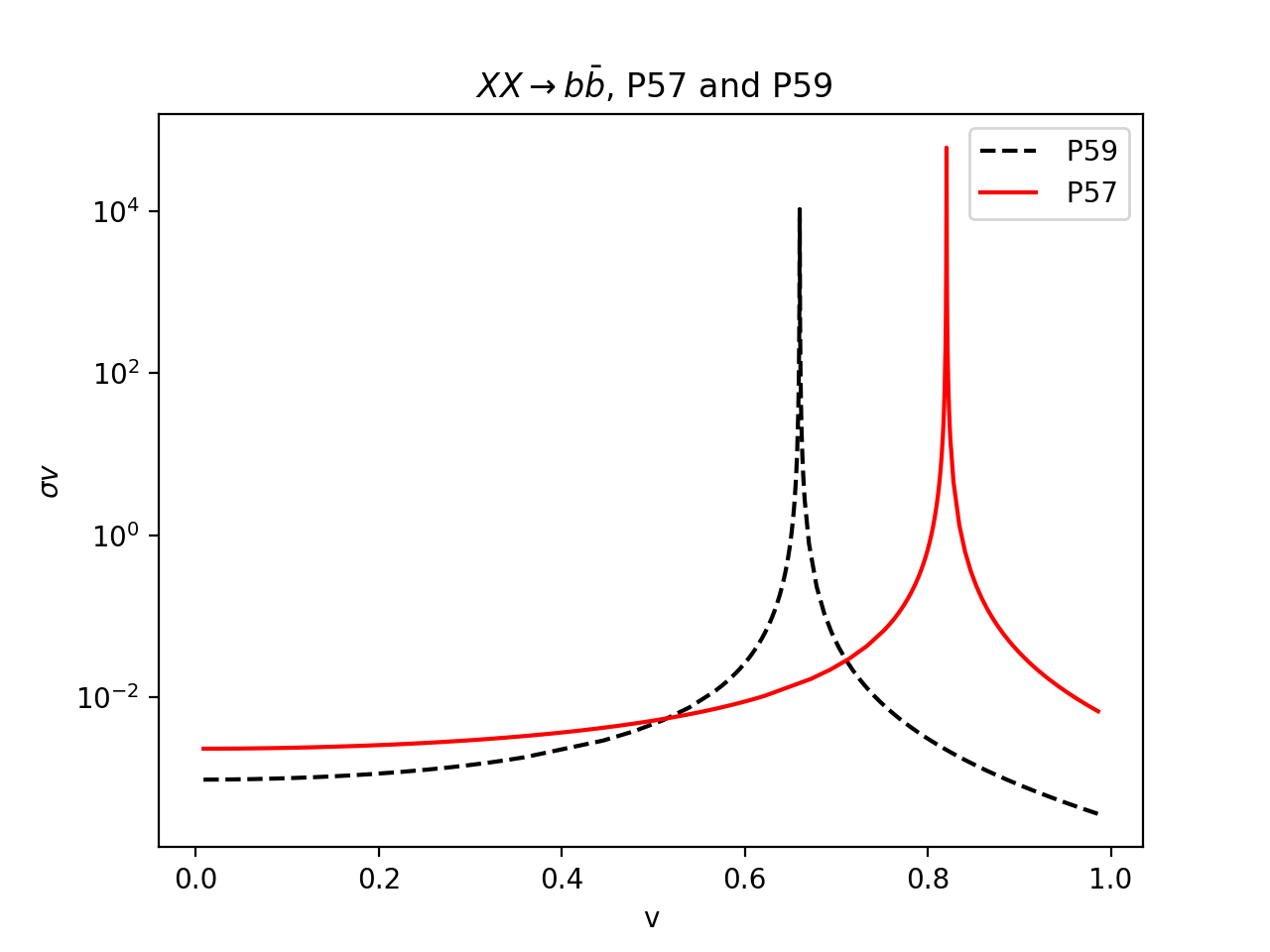}
\includegraphics[scale=0.50]{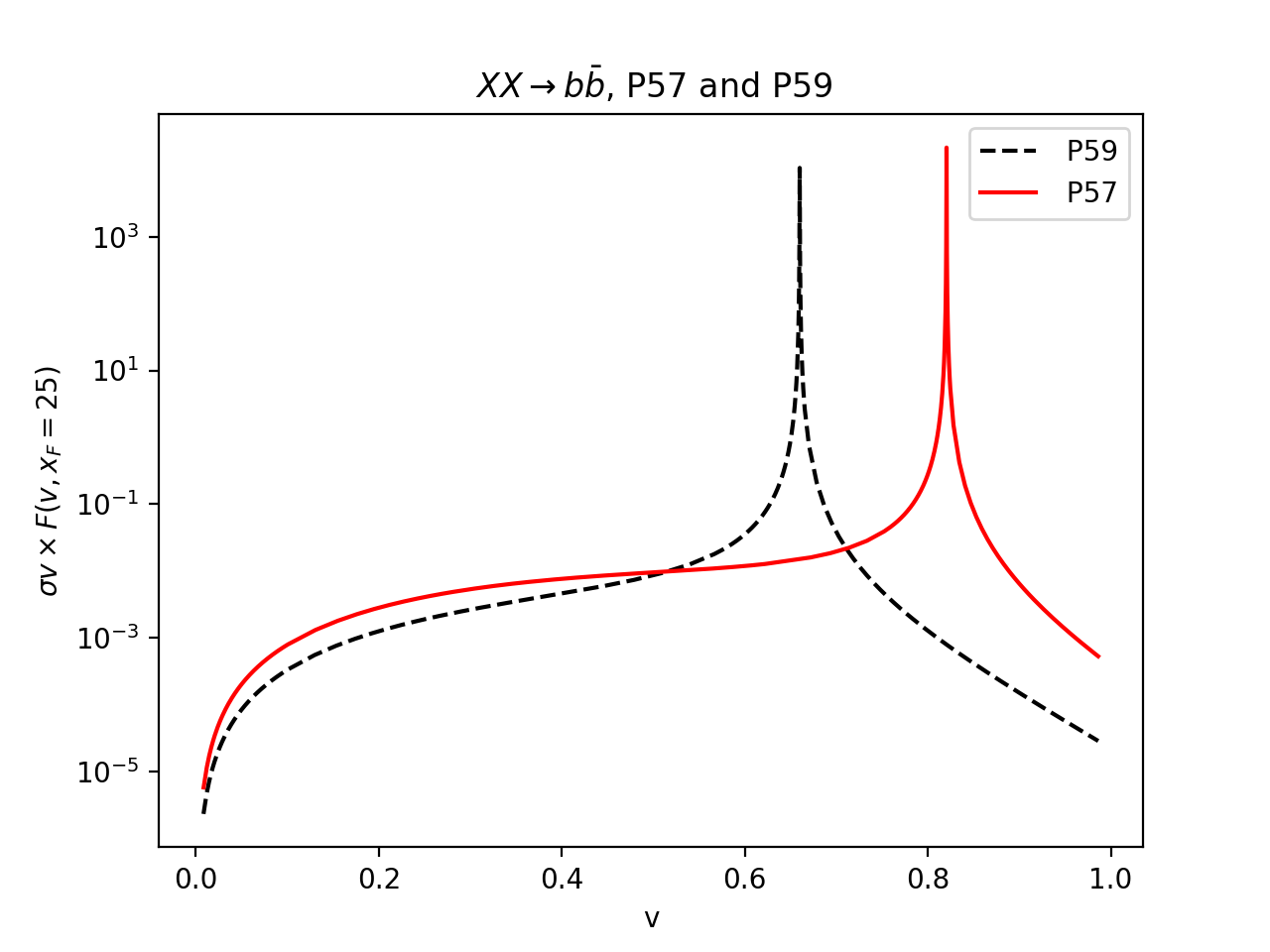}
\caption{\it  \label{fig:xxtobbtree-picstructure} The relative velocity dependence of the tree-level $XX \to b\bar b$ cross-section times $v$ for Points P57 and P59 in units of $10^{-26} {\rm cm}^3{\rm s}^{-1}$ (note the $\log$ scale). We also show the weighted cross-sections with the velocity distribution~\cite{OurPaper1_2020} with the freeze-out parameter $x_F=25$.}
\end{center}
\end{figure}

The annihilation cross-section and its velocity dependence is shown in Figure~\ref{fig:xxtobbtree-picstructure}. The peaks are extremely sharp because of the extremely narrow width of the SM Higgs with $\Gamma_h/M_h=3.26 \times 10^{-5}$.

We now understand that the P57 contribution to the relic density occurs at the edge of the velocity function, $v\sim 0.82$, in order to catch the Higgs resonance. It therefore needs a much larger value of $\l_L$ than P59 that captures the Higgs peak at $v\sim 0.66$. While converting the cross-sections into a contribution to the relic density, thermal averaging and convolution with the velocity distribution is performed. The weight of the velocity distribution is important. This convolution reduces somehow the weight of the resonance that occurs at large $v$ compared to the off resonance contributions that occur, say, at $v\sim 0.4$, see Figure~\ref{fig:xxtobbtree-picstructure}. 
Therefore, possible   changes around the peak and possible non-resonant loop-induced contributions  are important to study~\cite{OurPaper1_2020}. In any case, we need to very carefully scrutinise the peak. Another very important technical point is that we need to scan the Higgs resonance extremely carefully, both for the $b \bar b$ and the $ W W^\star$ channels. This is the reason why will be  generating a very large set of data points around the resonance. 


\section{The $b \bar b$ final state in the Higgs resonance region: one-loop electroweak corrections}
\label{sec:oneloopbb}
Electroweak loop calculations with width effects are very complicated, not only because of the numerical integration over the very narrow width region but also conceptually. First of all, the width is a resummation of a loop effect. Analyticity relates the width to the imaginary part of the self-energy of the corresponding particle. Introducing the width at the tree-level means that we are including a part of a higher order loop calculation (imaginary part of the two-point function). To be consistent and not double count contributions, at one-loop, the self-energy contribution (included already at tree-level as a part of $\Gamma_h$) must be subtracted. Moreover, the mixing of different orders in perturbation should be excised with care so as not to introduce a breaking of gauge invariance. In the case at hand, for the $b \bar b$ channel, the latter issue is easy to deal with.\\
Renormalisation is a coherent definition of the independent parameters of the model. Taking as input parameters all the masses of the scalars of the IDM, two extra independent parameters remain to be defined. The first one is $\l_L$. With a spectrum like the one in P57 and P59, $h \to XX$ is open. $\l_L$ is defined most easily (parametrically also) from $\Gamma (h \to XX)$ through Equation~\ref{eq:pwhinv}. We derive the associated counterterm in this full On-Shell (OS) scheme in~\cite{OurPaper1_2020}. Naturally, since $\Gamma(h\ra X X)$ is used as an input, it has the same value at all orders of perturbation. By construction, no loop correction affects the observable $\Gamma(h \ra XX)$. Another parameter, $\l_2$, describes the self-interaction between the scalars, {\it i.e., it resides solely within the scalar sector}. For the annihilation of DM into the SM particles, the cross-sections do not depend on $\l_2$ and therefore a counterterm for $\l_2$ for the annihilation cross-sections is not called for. However, $\l_2$ can indirectly affect the value of the cross-section, an effect akin to the top quark dependence of LEPII observables. 

\subsection{$XX \to b \bar b$ at one-loop: The theoretical set-up}
\begin{figure}[htbp]
\begin{center}
\includegraphics[scale=0.3]{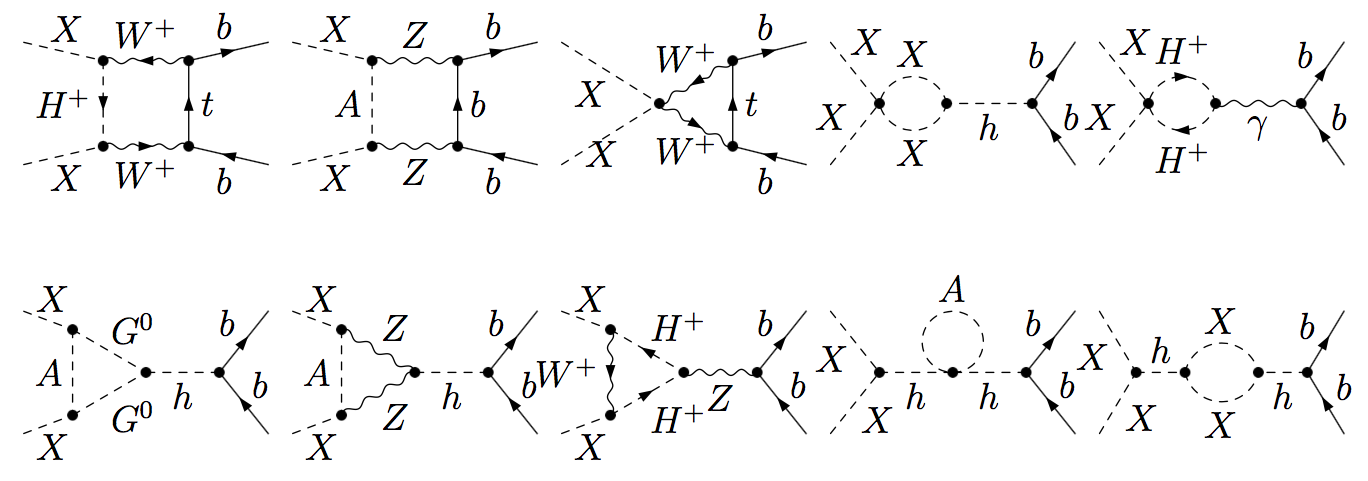}  
\caption{\it  \label{fig:xxtobboneloopfeynman} A selection of one-loop electroweak corrections diagrams contributing to $X X \to b \bar b$. The first 3 diagrams of the first row are genuine boxes which are non-resonant. The last 2 diagrams in the first row show examples where $\l_2$, rescattering in the dark sector, contributions can be induced at one-loop. The second row includes vertex corrections and self-energy contributions to the SM Higgs that can become resonant.}
\end{center}
\end{figure} 
While at tree-level the structure of the $XX \to b\bar b$ is simple since it proceeds through the unique contribution of  Higgs exchange, at one-loop a variety of contributions are at play, in particular non-resonant, non-Higgs exchange contributions, as shown by the selection in Figure~\ref{fig:xxtobboneloopfeynman}. The question is, how  can one then introduce the Higgs width to regulate the cross-section at $s=M_h^2$ in a coherent gauge invariant manner now that we have many contributions? 

Let us first start by reviewing the amplitude at tree-level. In our case it is not only a simple structure but it is also a trivially gauge invariant amplitude. Prior to the introduction of the width, the amplitude can be written as 
\beqn
\label{eq:hbbnowidth}
{\cal A}_h^{0}=\frac{{{\cal M}_{hXX}}\; {{\cal M}_{hbb}}}{s-M_h^2}, \quad {{\cal M}_{hXX}}={{\cal M}(h \to XX)} , \quad {{\cal M}_{hbb}}={{\cal M}(h \to b \bar b)}.
\eeqn

${{\cal M}(h \to XX)}$ and ${{\cal M}_{hbb}}={{\cal M}(h \to b \bar b)}$ are the respective amplitudes of the transitions $XX\to h$ and $h\to b \bar b$.
The pole structure requires the introduction of the width

\beqn
\label{eq:higgswidthfac}
{\cal A}_h^{0,w}=\frac{{{\cal M}_{hXX}}\; {{\cal M}_{hbb}}}{s-M_h^2+i\Gamma_h M_h}=\underbrace{\frac{s-M_h^2}{s-M_h^2+i\Gamma_h M_h }}_{F_h} {\cal A}_h^{0}, \quad F_h=\frac{s-M_h^2}{s-M_h^2+i\Gamma_h M_h }.
\eeqn

The last equality shows that the genuinely tree-level amplitude has been rescaled by a kinematics factor, $F_h$. Being a kinematical factor, it is gauge invariant. In what follows, the regulated amplitude is gauge invariant. In the code, for such a simple process, the introduction of the width is trivial. We either correct the amplitude of equation~\ref{eq:hbbnowidth} (or for that matter the cross-section) directly by adding the width to the Higgs propagator or, which amounts to exactly the same, by multiplying the non-width expression of equation~\ref{eq:hbbnowidth} by the rescaling factor, $F_h=F_h(s)$. It is important to observe that we are using a fixed width that corresponds to the observable total width or the physical width that can be perturbatively computed through an all-order calculation and is gauge invariant. In particular, it contains the dominant leading order that appears at one-loop in the self-energy of the Higgs. The tree-level amplitude therefore includes a contribution from the one-loop calculation. In turn, in performing the full one-loop correction, we therefore need to subtract this one-loop contribution since it is already put in the {\it width-improved} tree-level expression, in order to avoid double counting this specific contribution both in the width-improved tree-level cross-section and in the one-loop calculation. Note that the introduction of the width, at tree-level, can be interpreted as a correction that amounts to 
\beqn
\frac{-i \Gamma_h M_h}{s-M_h^2} {\cal A}_h^{0}.
\eeqn

At one-loop, the full correction consists of {\em i)} the self energy insertion, $\pihhs$, {\em ii)} vertex corrections and {\em iii)} box corrections, see figure~\ref{fig:xxtobboneloopfeynman}. The self-energy corrections, from the Higgs two-point function, and its counterterm, $\cpihhs$, add up to 
\beqn
\pihhs+\cpihhs&=&\pihhs+(s-M_h^2) \delta Z_h -\delta M_h^2+\frac{3 \delta T}{v}\nonumber \\
&=& \pihhs+(s-M_h^2) \delta Z_h -\hat{\delta} M_h^2, 
\eeqn
where $\delta T$ is the tadpole counterterm and $\delta Z_h$ is the SM Higgs wave function renormalisation, see~\cite{FawziPhysRepGrace}.

The  Higgs mass counterterm is defined from the \underline{real} part of the self-energy of the Higgs calculated at the Higgs mass 
\beqn
\hat{\delta} M_h^2={{\cal{R}}}e \pihhm.
\eeqn

At one-loop, prior to the introduction of the (total) width (in order to keep gauge invariance explicit), the total amplitude writes~\footnote{Strictly speaking, the counterterms $\delta {{\cal M}_{hbb}}$ and $\delta {{\cal M}_{hXX}}$ here refer to the counterterms of the parameters defining the corresponding vertices.}
\beqn
\delta {\cal A}_h^{1}=\underbrace{\frac{{{\cal M}_{hXX}}\; {{\cal M}_{hbb}}}{s-M_h^2}\; \left( -\frac{\pihhs+\cpihhs}{s-M_h^2}\right)}_{\text{Higgs self-energy}}+\underbrace{\Bigg(\frac{\delta {{\cal M}_{hXX}} {{\cal M}_{hbb}}+{{\cal M}_{hXX}} \delta {{\cal M}_{hbb}}}{s-M_h^2}\Bigg)}_{\text{vertex,Higgs exch.}}+\underbrace{\Box_{XXbb}.}_{\text{box corr., non-res. cont.}}
\eeqn
What we have collected in $\Box_{XXbb}$ includes not only box diagrams but also vertex ($Z,\gamma$) exchanges which are not Higgs-resonant, examples of which are shown in figure~\ref{fig:xxtobboneloopfeynman}. 

From the Higgs two-point function contribution, we have
\beqn
\pihhs+\cpihhs=i {\cal {I}}m \pihhm+ (s-M_h^2)\delta Z_h +(\pihhs-\pihhm).
\eeqn
$i {\cal {I}}m \pihhm$ is the one-loop term that should be subtracted for the one-loop contribution because part of it is included in ${\cal A}_h^{0,w}$ (the tree-level width inserted cross-section).
Indeed,
\beqn
{\cal{I}}m\pihhm=\Gamma_h^{0} M_h
\eeqn
$\Gamma_h^{0}$ is the tree-level calculated width of the Higgs. The easiest way to implement this subtraction is to define a modified  on-shell renormalisation for the Higgs mass such that

\beqn
\label{eq:improvedOS}
\hat{\delta} M_h^2= \pihhm, 
\eeqn

\noindent in other words the Higgs mass counterterm is defined from the full two-point function of the Higgs and not only from its real part. Then the imaginary part (width at tree-level) is subtracted automatically. Note that since $M_h^2 \propto \lambda_h v^2$, this means that we need to introduce imaginary counterterm for the Higgs self-coupling. Since the processes we study do not call for this coupling apart from the Higgs mass, this is not required for this simple process. 

The one-loop contribution (before regularisation with the total width) is 
\beqn
\label{eq:resbbh11}
\delta {\cal A}_h^{1}&=&\frac{{{\cal M}_{hXX}}\; {{\cal M}_{hbb}}}{s-M_h^2}\; \left( -\delta Z_h-\frac{\pihhs-\pihhm}{s-M_h^2} \right) \nonumber \\
&+&\Bigg(\frac{\delta {{\cal M}_{hXX}} {{\cal M}_{hbb}}+{{\cal M}_{hXX}} \delta {{\cal M}_{hbb}}}{s-M_h^2}\Bigg)+\Box_{XXbb}.
\eeqn

Maintaining gauge invariance of the above amplitude while introducing a Higgs width, our scheme consists of \underline{applying the factorising factor, $F_h$, to the {\bf whole} amplitude of Equation~\ref{eq:resbbh11}}.

The regularised one-loop corrected, gauge invariant, amplitude is then 
\beqn
\label{eq:Xbbone-loop-factor}
\delta {\cal A}_h^{1,w}&=&\frac{s-M_h^2}{s-M_h^2+i\Gamma_h M_h}\delta {\cal A}_h^{1} \nonumber \\
&=&\frac{{{\cal M}_{hXX}}\; {{\cal M}_{hbb}}}{s-M_h^2+i\Gamma_h M_h}\; \left( -\delta Z_h-\frac{\pihhs-\pihhm}{s-M_h^2} \right) \nonumber \\
&+&\Bigg(\frac{\delta {{\cal M}_{hXX}} {{\cal M}_{hbb}}+{{\cal M}_{hXX}} \delta {{\cal M}_{hbb}}}{s-M_h^2+i\Gamma_h M_h}\Bigg)\nonumber \\
&+& \frac{s-M_h^2}{s-M_h^2+i\Gamma_h M_h}\Box_{XXbb}.
\eeqn
The drawback of the scheme is that the non-resonant contributions gathered in $\Box_{XXbb}$ vanish at the resonance, $s=M_h^2$. By default, in the code we first calculate the one-loop amplitude without the width within the {\it modified} OS scheme (Equation~\ref{eq:improvedOS}) to avoid double counting. In a second step, we apply the $F_h$ factor to the full amplitude to obtain Equation~\ref{eq:Xbbone-loop-factor}. Observe that the factor containing $\delta Z_h$ in Equation~\ref{eq:Xbbone-loop-factor} (first line) also has a term with a denominator $(s-M_h^2)$. For this term, no width can be introduced but the expression is fully regularised because of the difference appearing in the accompanying numerator.\\

\noi As promised, let us get back to equation~\ref{eq:treeXXbb} and see how it compares with a full one-loop calculation. Since we have implemented an OS shell scheme where $\Gamma(h\to XX)$ is used in {\it lieu} of $\l_L$ and where we are making use of the total width $\Gamma_h$ as input, at the peak $s=M_h^2$, $\Gamma(h\to XX)$ is unchanged but we expect a very small electroweak correction to $\Gamma(h \to b\bar b)$, such that  $\Gamma(h \to b\bar b) \equiv \Gamma(h \to b\bar b)^{\text{tree}} \to \Gamma(h \to b\bar b)^{\text{1-loop, EW}}$(which, as we argued before is insensitive to the dark sector at one-loop). But, more importantly, loop effects bring in new contributions that are not resonant and that are not described by the Breit-Wigner {\it approximation}. But, for the sake of comparison, at $s \simeq M_h^2$, we can seek an improved tree-level electroweak cross-section by an effective rescaling of $\Gamma(h \to b \bar b)^{\text{1-loop, EW}}$. This is a very small correction that we have calculated to be
\beqn
\label{eq:hbbewrc}
\frac{\Gamma(h \to b \bar b)^{\text{1-loop, EW}}}{\Gamma(h \to b\bar b)}=1.0166.
\eeqn
Hence the {\em improved} $b\bar b$ cross-section (that will miss one-loop non-resonant corrections) writes as
\beqn
\label{eq:treeXXbbimp}
\sigma^{\text{imp.}}{(X X \stackrel{h}{\longrightarrow} b \bar b})\; v=
\frac{ 64 \; \pi \frac{\Gamma(h\ra X X)}{\sqrt{1-4M_X^2/M_h^2}} \Gamma(h\ra b\bar{b})^{\text{1-loop, EW}}}{(s -M_h^2)^2+
\Gamma_h^2 M_h^2}.
\eeqn

\subsection{Results for $XX \to b \bar b$ at one-loop}
\begin{figure}[hbtp]
\begin{center}
\includegraphics[width=0.45\textwidth]{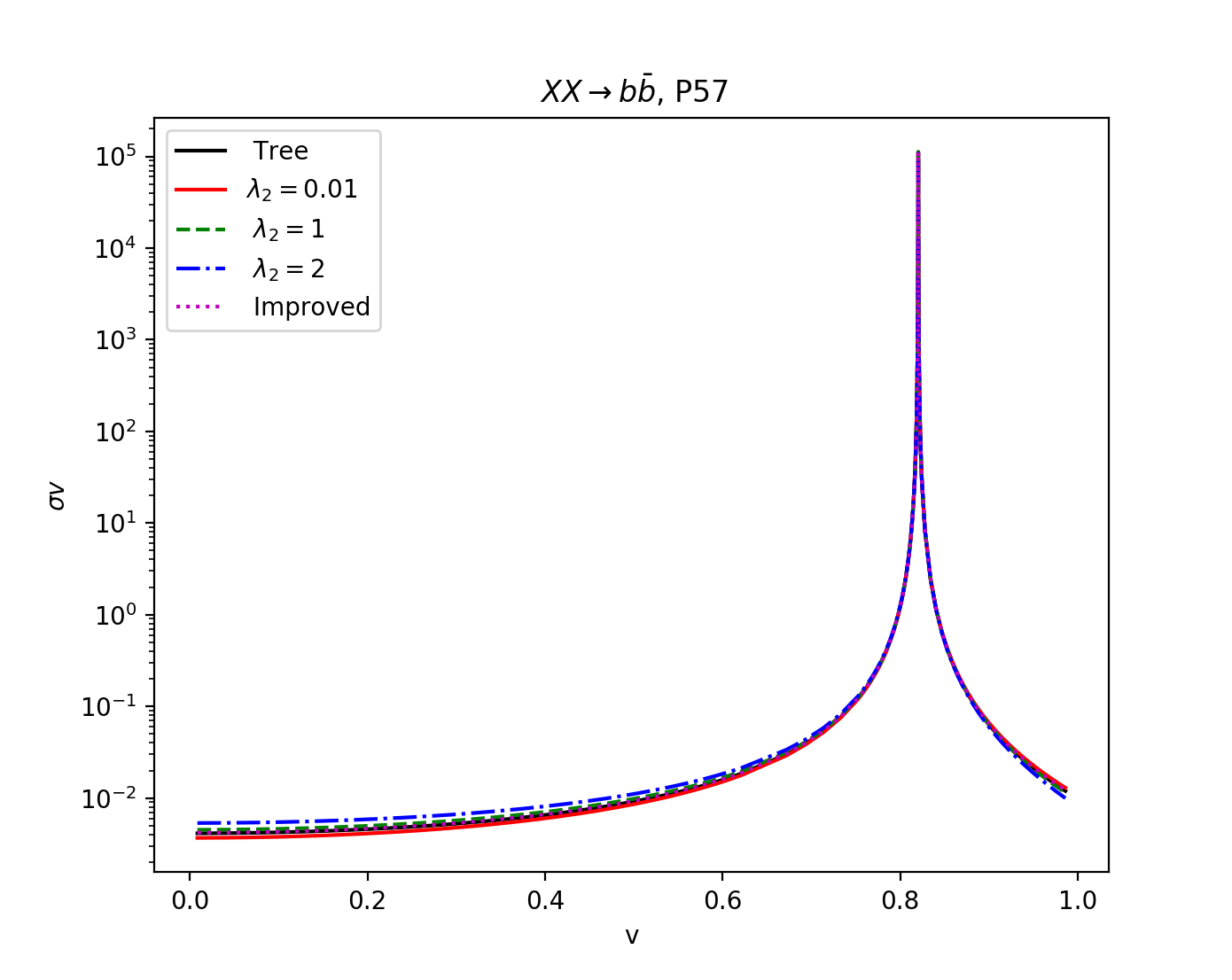}
\includegraphics[width=0.45\textwidth]{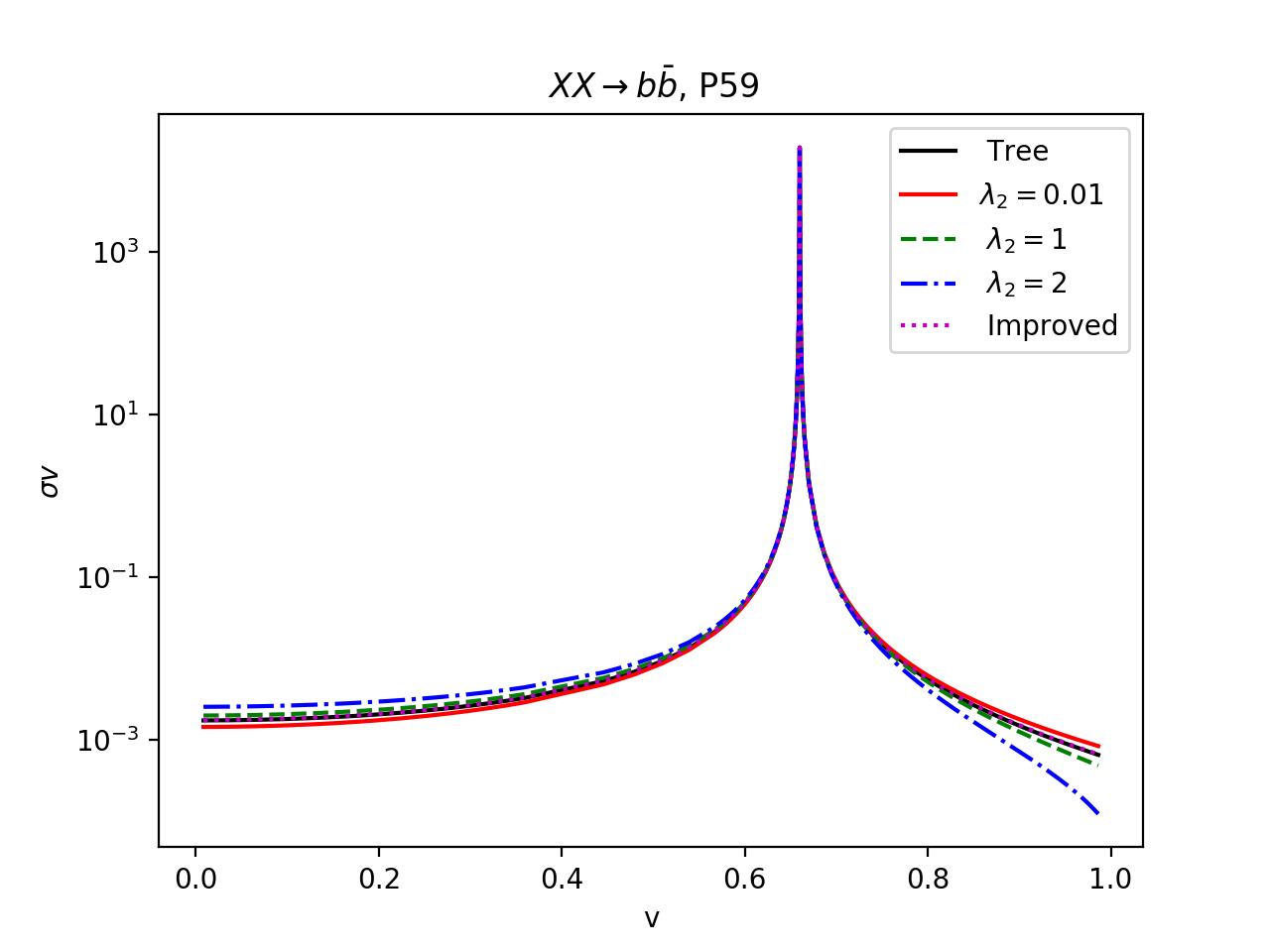}
\includegraphics[width=0.45\textwidth]{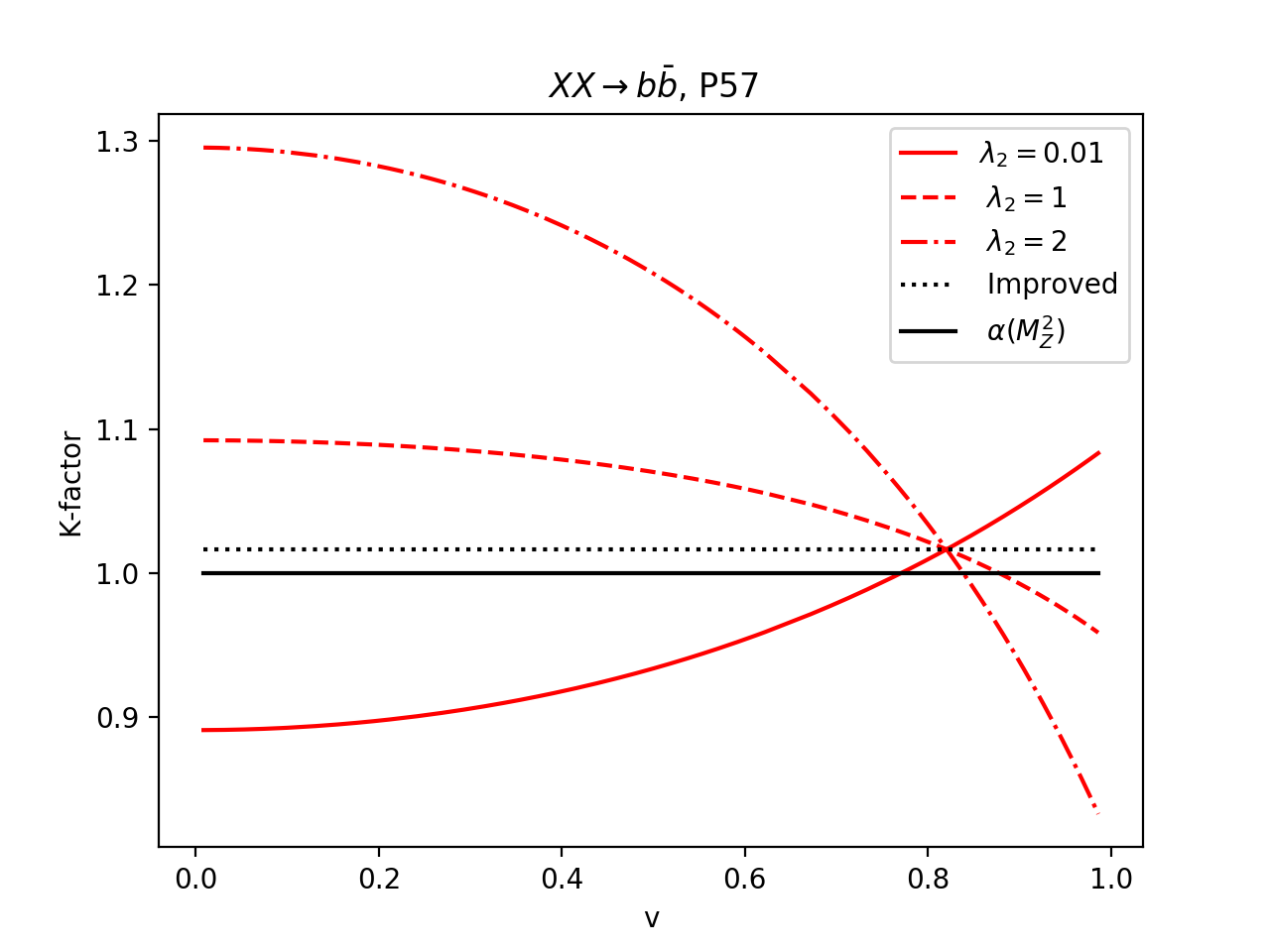}
\includegraphics[width=0.45\textwidth]{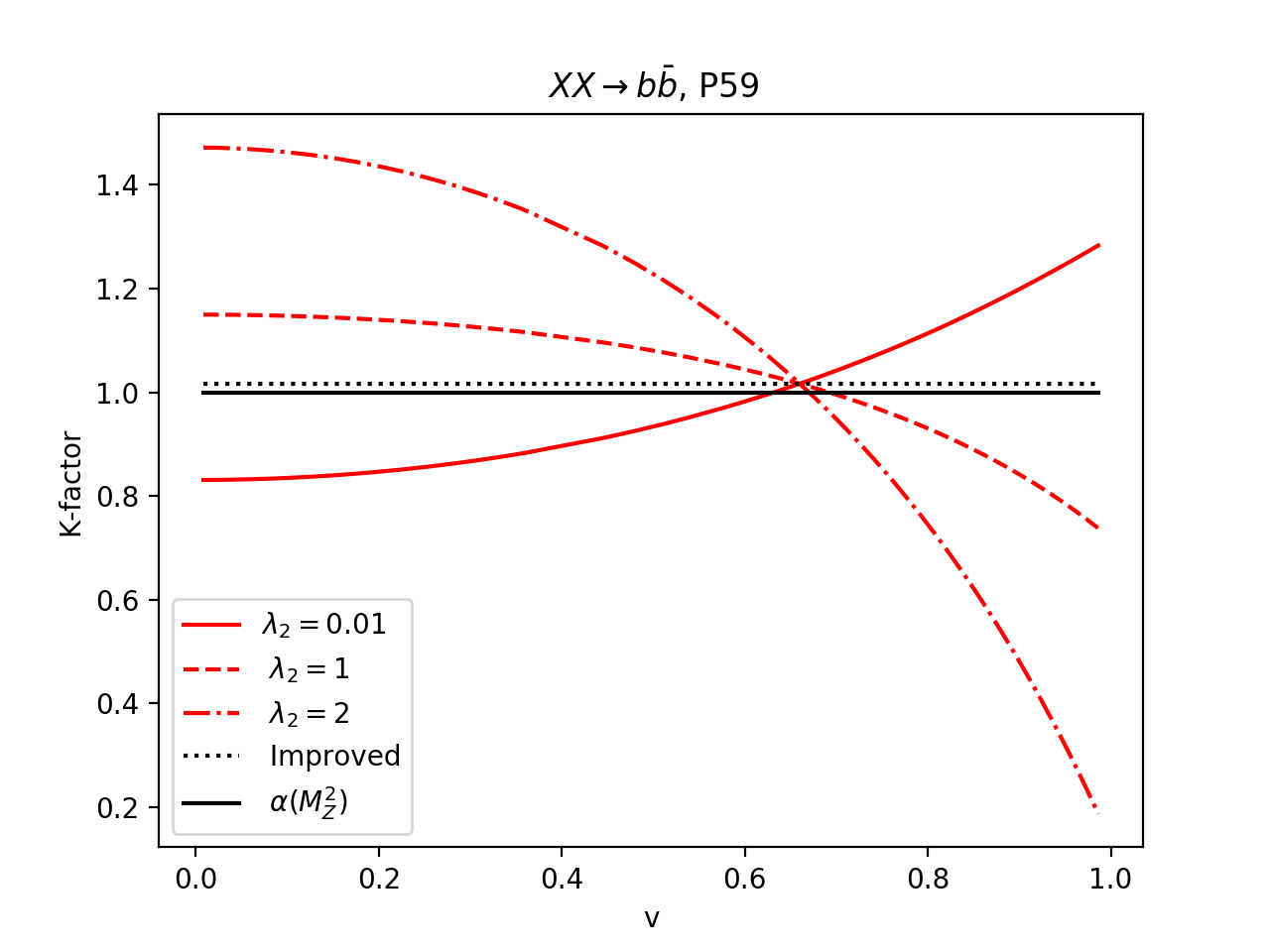}
\includegraphics[width=0.45\textwidth]{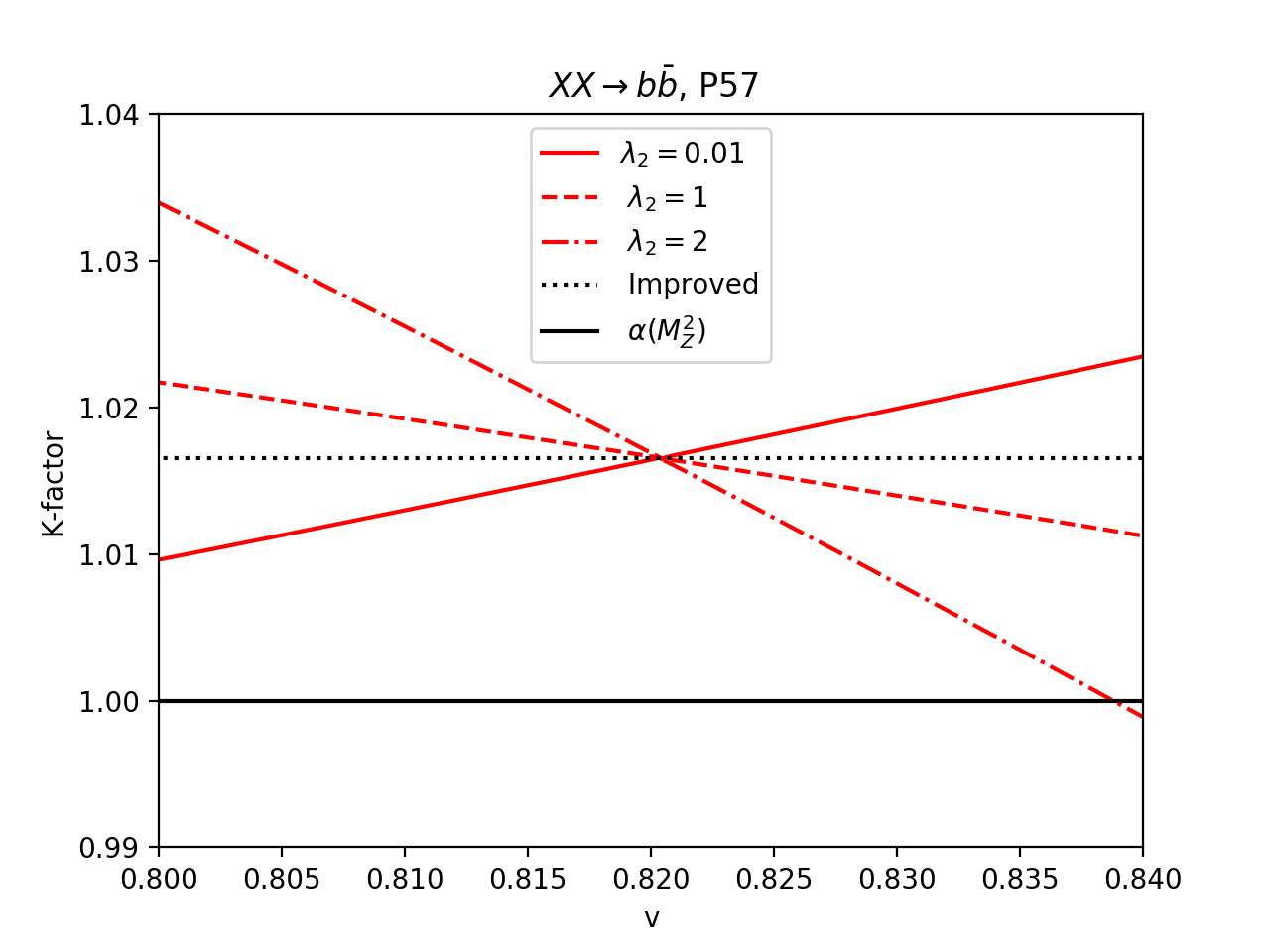}
\includegraphics[width=0.45\textwidth]{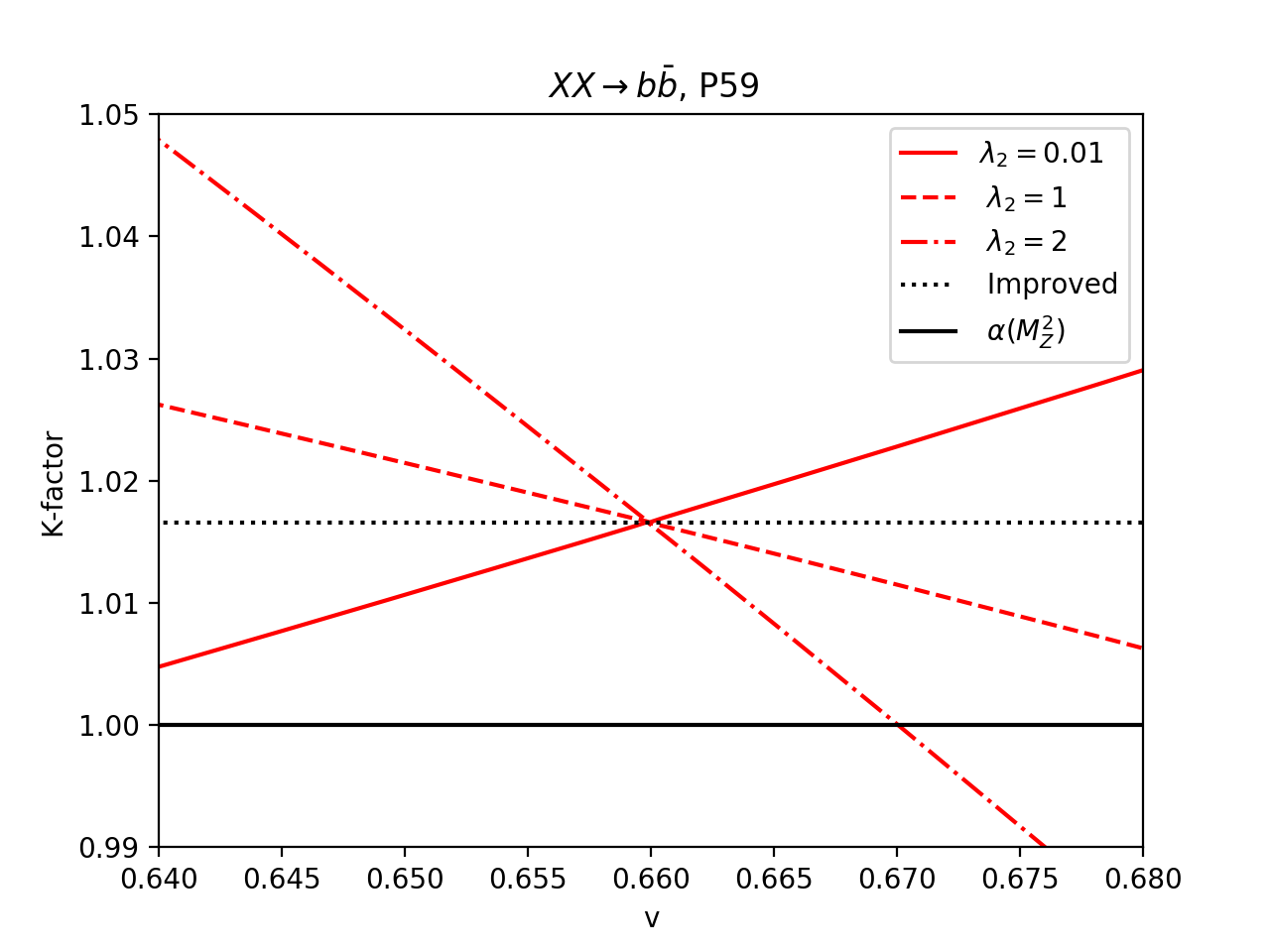}
\caption{\it  \label{fig:xxtobbloopresults} The relative velocity dependence of the one-loop $XX \to b\bar b$ cross-section times $v$ for Points P57 (left panels) and P59 (right panels) in units of $10^{-26} {\rm cm}^3{\rm s}^{-1}$ (note the $\log$ scale). The lower panels help better see the effect of the loop corrections by giving the relative corrections (with respect to the tree-level) over the full velocity range as well as a zoom around the peak. The $\l_2$ dependence of the one-loop results is given. Also shown is the result of using the full electroweak correction to $\Gamma(h \to b \bar b)$ (the improved cross-section given by Equation~\ref{eq:treeXXbbimp}). We also give the result of using, at tree-level, $\alpha(M_Z^2)$ instead of $\alpha$.}
\end{center}
\end{figure}
Our numerical results for the full one-loop electroweak corrections for P57 and P59 are shown in figure~\ref{fig:xxtobbloopresults}. Apart from the location of the Higgs peak in terms of the relative velocity, the results for both benchmarks are very similar. On the whole, the dependence of the velocity and the peak structure are maintained when we move from the tree-level to the one-loop corrected results. Because of the very narrow-peaked structure, the effect of the radiative corrections at the level of the cross-sections is hardly distinguishable. To see the subtle details, which amount nonetheless to important corrections, we look at the normalised corrections with respect to the tree-level results. First, the numerical computation of the tree-level cross-section based on the use of $\alpha(M_Z^2)$ confirms that we obtain exactly the same results, for all $v$, as the use of $\alpha(0)$ (the ratio is equal to $1$ throughout all $v$). This quantitatively substantiates the remark we made earlier that the $b \bar b$ cross-section is insensitive to $\alpha$. Away from $s=M_h^2$, the $\l_2$ dependence of the cross-section is important and grows as we further move away from the peak. The $\l_2$ dependence, though qualitatively similar between the two models, is rather different numerically. For P57, the difference in the relative correction between $\l_2=0.01$ and $\l_2=2$ reaches $40\%$ while for P59 this difference amounts to more than $60\%$. As expected, at exactly the Higgs resonance, our computation confirms that the $\l_2$ dependence drops out. This is very understandable since the renormalisation of $\l_L$ is based on the OS scheme where $\Gamma(h \to XX)$ is used as input, so any $\l_2$ dependence has been absorbed in the OS value of $\l_L$ at $s=M_h^2$. Moreover, the value of the correction at the resonance is reproduced exactly by the use of $\Gamma(h  \to b\bar b)^{\text{1-loop, EW}}$ rather than the pure tree-level which defines what we termed the improved cross-section. It rests that within $2\Gamma_h$ from the peak at $M_h$, the electroweak correction to the this annihilation channel is of the order of $2\%$. The most important lesson is that this small relative correction is in large part due to the OS scheme we have used that takes $\Gamma(h \to XX)$ as input and that the implementation of the width is such that all one-loop induced contributions are multiplied by $s-M_h^2$, which leaves then only the Higgs resonance. The remaining small correction, which is a result of the electroweak correction $\Gamma(h\to b \bar b)$, is very small. Important corrections occur away from the resonance. The interesting observation is that they reveal the indirect $\l_2$ contribution hidden in the dark sector. 


\section{$X X \stackrel{h}{\longrightarrow} W W^\star$ at tree-level}
\label{sec:treehWW}
\begin{figure}[htbp]
\begin{center}
\includegraphics[scale=0.35]{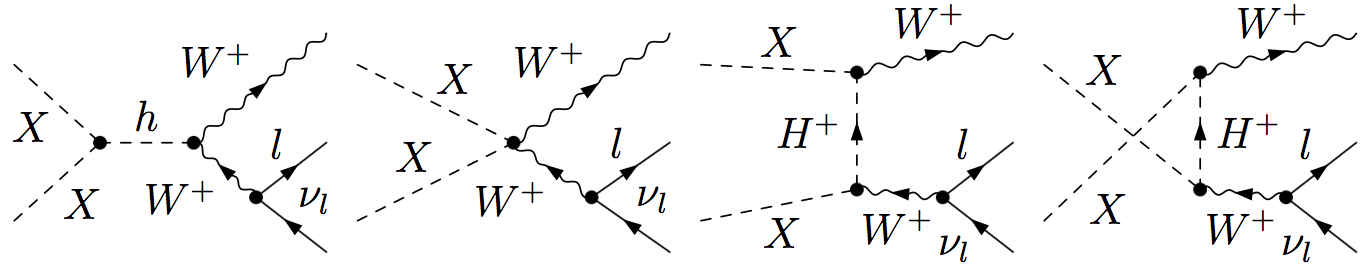}
\caption{\it  \label{fig:HresWlnl} Tree-level diagrams contributing to $XX \to W l \nu_l$. Apart from the Higgs exchange contribution with $h^\star \to W l \nu_l$ there are other non-Higgs resonant contributions. There are other diagrams with $h \to l l^\star \to l W \nu_l$ that gives effects proportional to the lepton mass which we do not show here but which are included in our full calculation.}
\label{fig:FeynXXbbtree}
\end{center}
\end{figure}
\begin{figure}[htbp]
\begin{center}
\includegraphics[scale=0.4]{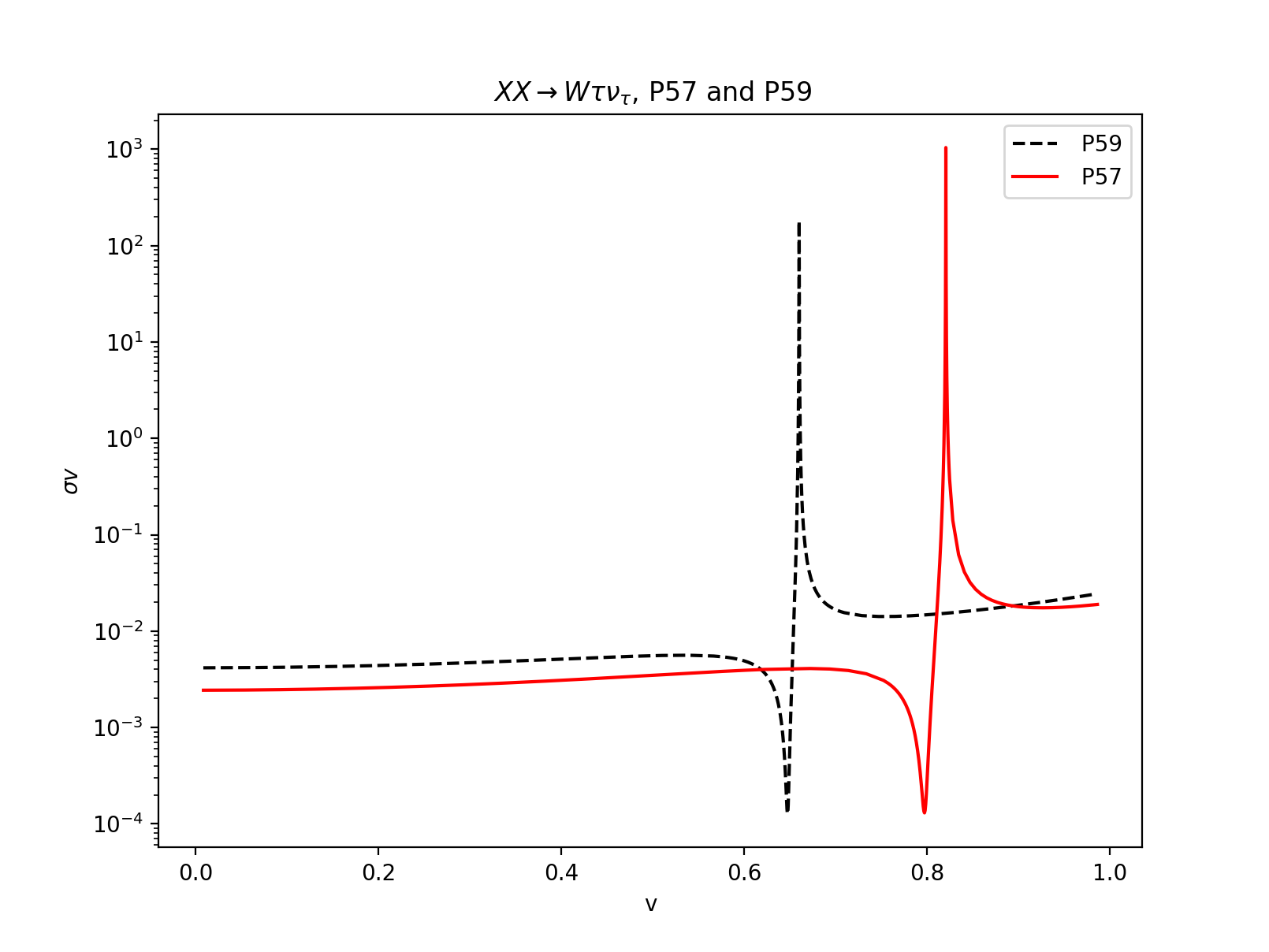}
\includegraphics[scale=0.4]{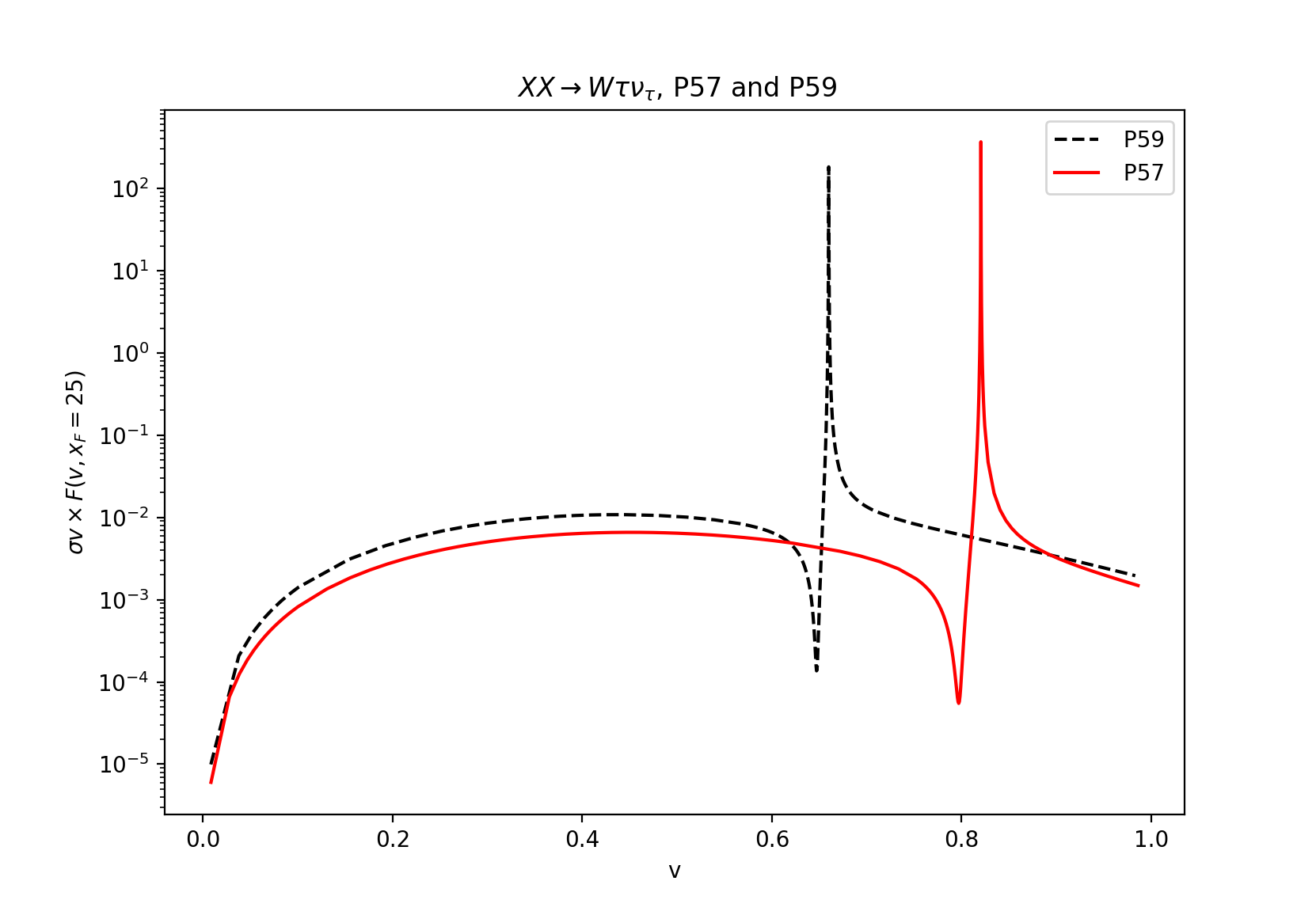}
\caption{\it  \label{fig:xxtowenetree-picstructure} As in Figure~\ref{fig:xxtobbtree-picstructure} but for $XX \to W \tau \nu_\tau$. Note the behaviour of the peak structure that is a characteristic of an interference of a smooth contribution with a resonance. The panel on the right represents the convolution of the cross-sections with the velocity distribution with the same parameters as in Figure~\ref{fig:xxtobbtree-picstructure}.}
\end{center}
\end{figure}

 
What we call $XX \to W W^\star$ is in fact $XX \to W f \bar f^\prime$, a $2 \to 3$ amplitude. These cross-sections are studied in Ref.~\cite{OurPaper3_2020} but in the case where the DM mass was large enough so that the SM Higgs resonance is not reached and there was no need to include the Higgs width. In~\cite{OurPaper2_2020}, $XX \to W W^\star$ contributes subdominantly to the relic density and while the mass spectrum is such that the Higgs resonance is hit, we have $\l_L=0$ so that the Higgs exchange is absent at tree-level and at one-loop and the induced Higgs exchange crosses the resonances smoothly. The scenarios in the present paper is that the SM Higgs resonance is crossed with $\l_L\neq 0$. The inclusion of the Higgs resonance must be implemented both at tree-level and at one-loop. The issue is more involved than in the case of the annihilation into $b \bar b$ we just studied. Here, the total contribution involves a resonant Higgs exchange contribution {\underline{and}} a gauged, non-resonant, amplitude. Moreover, the $\l_L$ dependence of the $XX \to W W$ cross-section stems not only from the pure SM Higgs (resonant) contribution but also from the quartic coupling $XXWW$. Converting (or regularising) only one of these contributions (the Higgs exchange) to include the Higgs width, the subtle $\l_L$ dependence is broken in the sense that we would be providing a width for the Higgs for only a part of the $\l_L$ dependent amplitude. This is akin to a breaking of gauge invariance when only a part of the amplitude is endowed with a width dependent propagator. One needs to be careful at how the width is introduced to regulate the {\underline{full}} amplitude. \\
\noi Here again it is best to start from the full zero-width tree-level amplitude. This amplitude can be written in terms of its pole structure as a Laurent series in $1/(s-M_h^2)$. The full amplitude at tree-level can then be written as 
\beqn
\label{XXWWstar-higgsres0}
{\cal {A}}_{XXWW}^{0}={\cal {A}}_{XXWW}(g^2,\l_L)+ \frac{  {{\cal M}_{hXX}}\;{{\cal M}_{hWff}}  }{(s-M_h^2)},
\eeqn 
where ${{\cal M}_{hXX}}\sim \l_L$ and ${{\cal M}_{hWff}}$ represent, respectively, the transition amplitude of the SM Higgs to the DM particles ($h \to XX$) and the transition amplitude, $h \to Wf\bar{f^\prime}$.\\ 
\noi As we did with the $b \bar b$ channel, we introduce the width at tree-level through the overall kinematical fudge factor $F_h$ (see Equation~\ref{eq:higgswidthfac}) to the {\underline{full}} amplitude, such that 
\beqn
\label{XXWWstar-higgsres1}
{\cal {A}}_{XXWW}^{0,w}& =&F_h\Bigg({\cal {A}}_{XXWW}(g^2,\l_L)+ \frac{{{\cal M}_{hXX}}\;{{\cal M}_{hWff}}}{(s-M_h^2)}\Bigg)\nonumber \\
&=&
\frac{(s-M_h^2)}{(s-M_h^2)+i\Gamma_h M_h}{\cal {A}}_{XXWW}(g^2,\l_L) + \frac{{{\cal M}_{hXX}}\;{{\cal M}_{hWff}}}{(s-M_h^2)+i\Gamma_h M_h}.
\eeqn 
This treatment is the same as what we apply at tree and one-loop levels for $XX \to b \bar b$. Again, at $s=M_h^2$, the non resonant contribution is effectively set to zero with this procedure. The ensuing cross-section can be parametrised in terms of this pole structure as 

\beqn
\label{eq:CSXXWWstartree}
\sigma{(X X \to W f^\prime \bar f})^{0,w}&=&
 \frac{\; 64 \; \pi \frac{\Gamma(h\ra X X)}{\sqrt{1-4M_X^2/M_h^2}} \Gamma(h\ra W f^\prime \bar f)}{(s -M_h^2)^2+
\Gamma_h^2 M_h^2} +F_h^2 \sigma_{XXWW}(g^2,\l_L) \nonumber \\ 
&+& \frac{s-M_h^2}{(s-M_h^2)^2+\Gamma_h^2 M_h^2}\text{Ps}{\cal{R}}e\Bigg({\cal {A}}_{XXWW}(g^2,\l_L) ({{\cal M}_{hXX}}\;{{\cal M}_{hWff}}) ^*\Bigg),
\eeqn
where Ps represents the phase space factor. This is the tree-level cross-section provided by the code. In the above parametrisation, in the first term of Equation~\ref{eq:CSXXWWstartree}, the width $\Gamma(h\ra W f^\prime \bar f)$ should represent the tree-level value, a result similar to what was presented for the $b \bar b$ channel. Figure~\ref{fig:xxtowenetree-picstructure} shows the cross-section as a function of the relative velocity confirming the Higgs resonance and also the effect of the interference leading to an anti-resonance structure, with a dip followed by a spike at exactly the resonance around a very small region of $v(s)$. 

In accordance with the analysis for the $b \bar b$ final state, we anticipate and suggest an {\em improved} tree-level cross-section obtained as 
\beqn
\label{CSXXWWstarimproved}
\sigma{(X X \to W f^\prime \bar f})^{\text{improved}}&=&\frac{\Gamma(h\ra W f^\prime \bar f)_\text{1-loop}}{\Gamma(h\ra W f^\prime \bar f)_\text{tree}}
\sigma{(X X \to W f^\prime \bar f})^{0,w},  \\
\eeqn
so that the resonant contribution at the peak contains the full one-loop correction. This constant factor applied to the full tree-level cross-section will not change the behaviour of the velocity ($s$) dependence from what is shown in Figure~\ref{fig:xxtowenetree-picstructure}. We calculate

\beqn
\frac{\Gamma(h\ra W l \nu_l)_\text{1-loop}}{\Gamma(h\ra W l \nu_l)_\text{tree}} \simeq \frac{\Gamma(h\ra W q \bar q^\prime )_\text{1-loop}}{\Gamma(h\ra W q \bar q^\prime)_\text{tree}}  \simeq 1.087.
\eeqn

This result on the relative correction to the partial decay width is, to a precision of $0.02\%$, flavour blind and stems from the fact that the electroweak corrections to $W \to f \bar f^\prime$ are practically flavour blind~\cite{Bardin:1986fi, OurPaper3_2020}. The $8.7\%$ {\it relative} correction is the combination of the  full one-loop correction to $W\to f \bar f^\prime$ ($2.4\%$) and an effective correction at the level of the $hWW$ vertex ($6.3\%$) that is absorbed by using $\alpha(M_Z^2)$. This observation is important because it points to the fact that the full electroweak correction in these channels should be flavour blind and that the relative correction over the whole $v$ range is also flavour blind, a property that we already found in our study for this process away from the Higgs resonance~\cite{OurPaper2_2020, OurPaper3_2020}. We directly verify this property in this case. It is yet another verification of our automatic calculation and the performance of our code~\footnote{For massless fermions, the partial width of the charged $W$, including QCD corrections, can be written as $\Gamma(W \to q \bar q^\prime)=N_C \; \frac{\alpha M_W}{ 12 (1-M_W^2/M_Z^2)} |V_{qq^\prime}|^2 (1+\frac{3}{8} \frac{N_C^2-1}{N_C} \frac{\alpha_s}{\pi})$, where $V_{qq^\prime}$ is the element of the KM matrix, and $\alpha_s$ is the QCD fine structure constant.}. 


\section{$XX \stackrel{h}{\longrightarrow} W f^\prime \bar f$ at one-loop}
\label{sec:loophWW}
The one-loop electroweak corrections to this $2\to 3$ process are challenging especially because we have to deal with the Higgs resonance. A large number of one-loop topologies are involved. A subset of these topologies can be found in Ref.~\cite{OurPaper2_2020, OurPaper3_2020}. The set up of the one-loop calculation with the insertion of the width goes through the same steps as those that we presented in the case of $b \bar b$. In a nutshell, the width is introduced through the $F_h$ factor for the entire amplitude. Therefore, we can write for the correction 
\beqn
\delta {\cal A}_{h_{WWff}}^{1,W}
&=&\frac{{{\cal M}_{hXX}}\; {{\cal M}_{hWff}}}{s-M_h^2+i\Gamma_h M_h}\; \left( -\delta Z_h-\frac{\pihhs-\pihhm}{s-M_h^2} \right) \nonumber \\
&+&\Bigg(\frac{\delta {{\cal M}_{hXX}} {{\cal M}_{hWff}}+{{\cal M}_{hXX}} \delta {{\cal M}_{hWff}}  }{s-M_h^2+i\Gamma_h M_h}\Bigg)\nonumber \\
&+& \frac{s-M_h^2}{s-M_h^2+i\Gamma_h M_h}\Box_{XXWWf}^{(5)}. \eeqn

$\delta {{\cal M}_{hWff}}$ pertains to the correction on the final state of the Higgs decay. It includes the vertex correction to $hW W^\star$, 2-point function for the $W$ propagator and the $4$ point function. $\Box_{XXWWf}^{(5)}$ relates to corrections that connect initial/final states which are 5-point function corrections. These contributions are not factorisable, with our implementation of the width, these contributions are effectively put to zero at exactly the Higgs resonance. 

\begin{figure}[hbtp]
\begin{center}
\includegraphics[width=0.45\textwidth,height=0.35\textwidth]{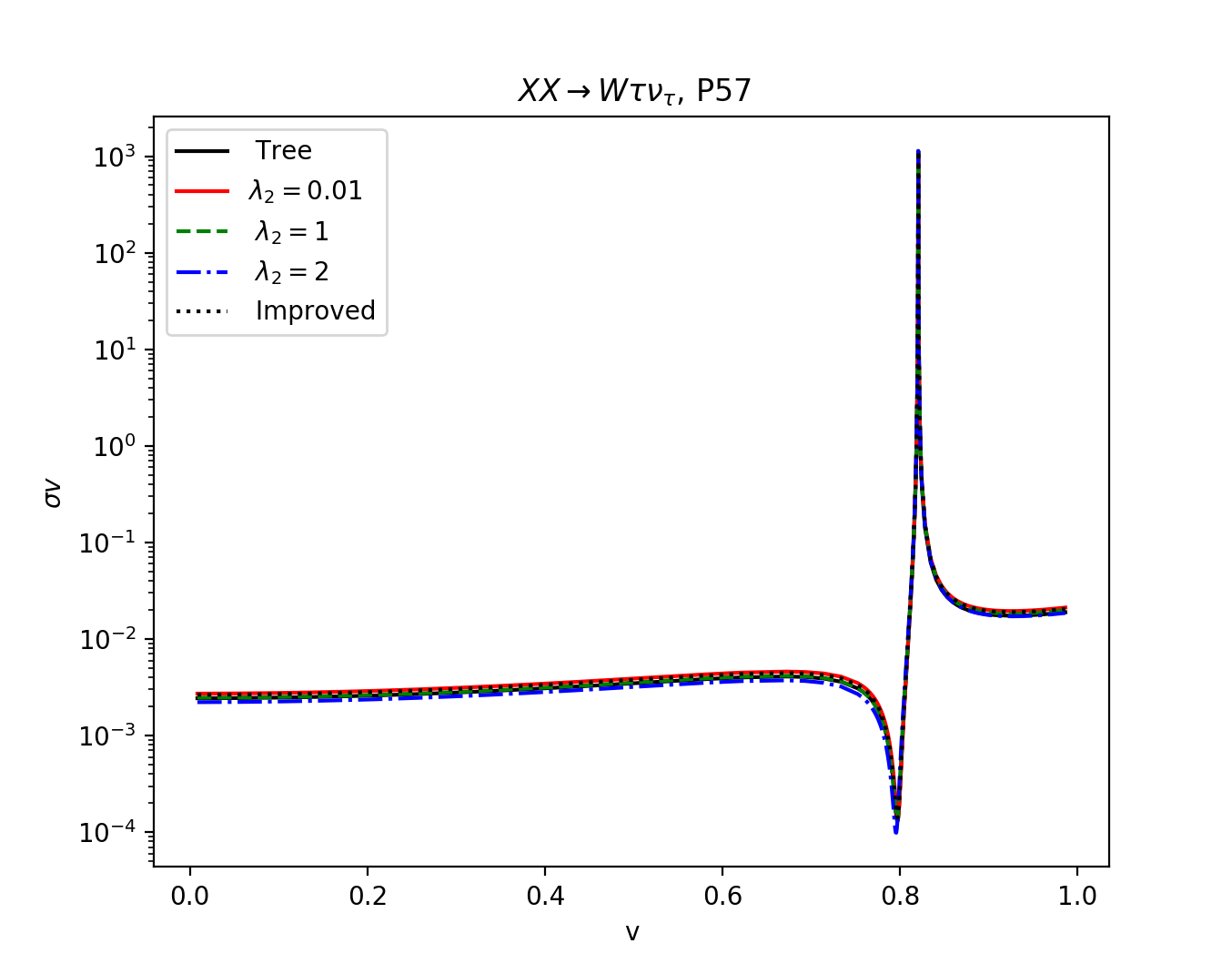}
\includegraphics[width=0.45\textwidth,height=0.35\textwidth]{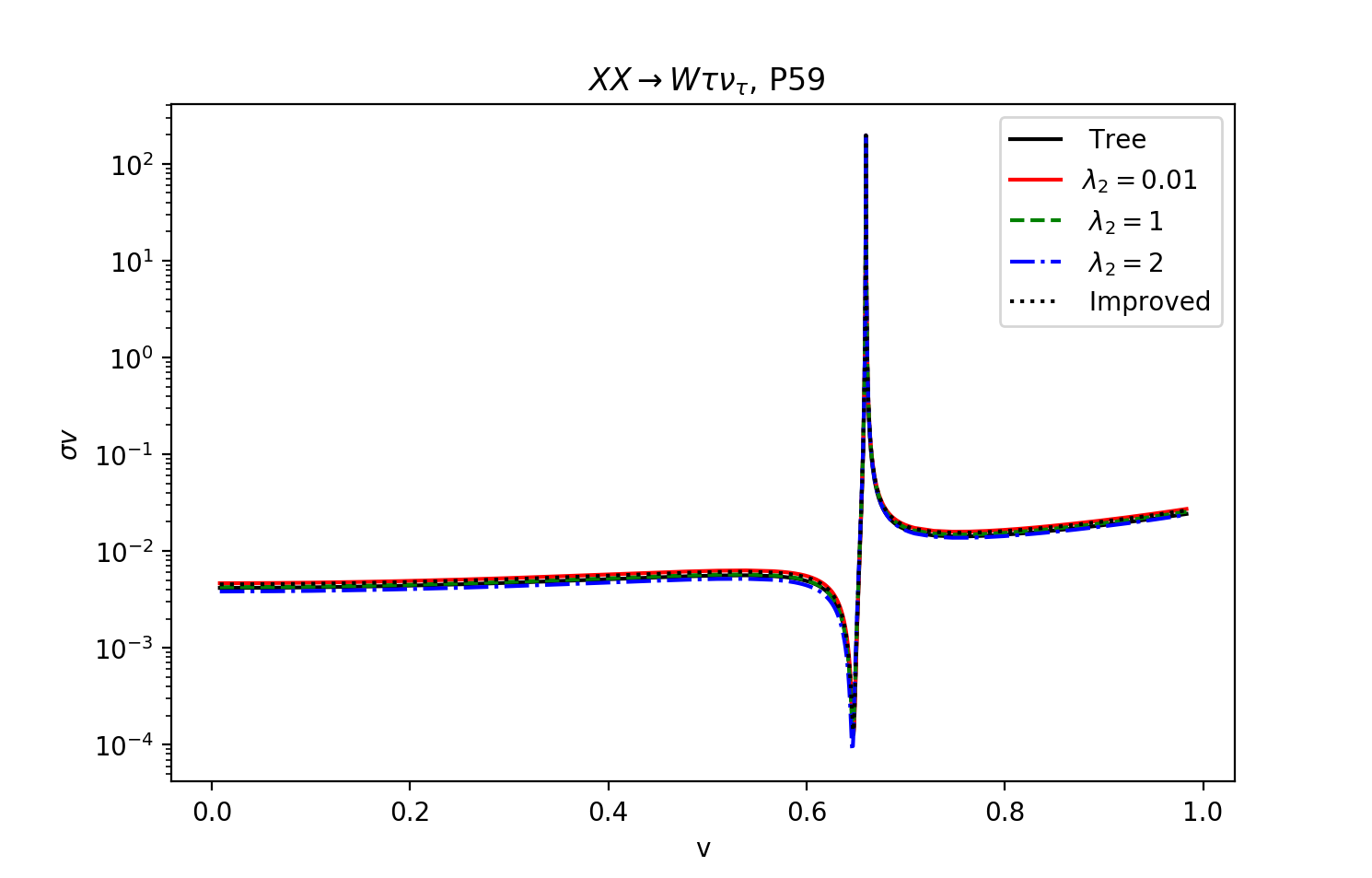}
\includegraphics[width=0.45\textwidth]{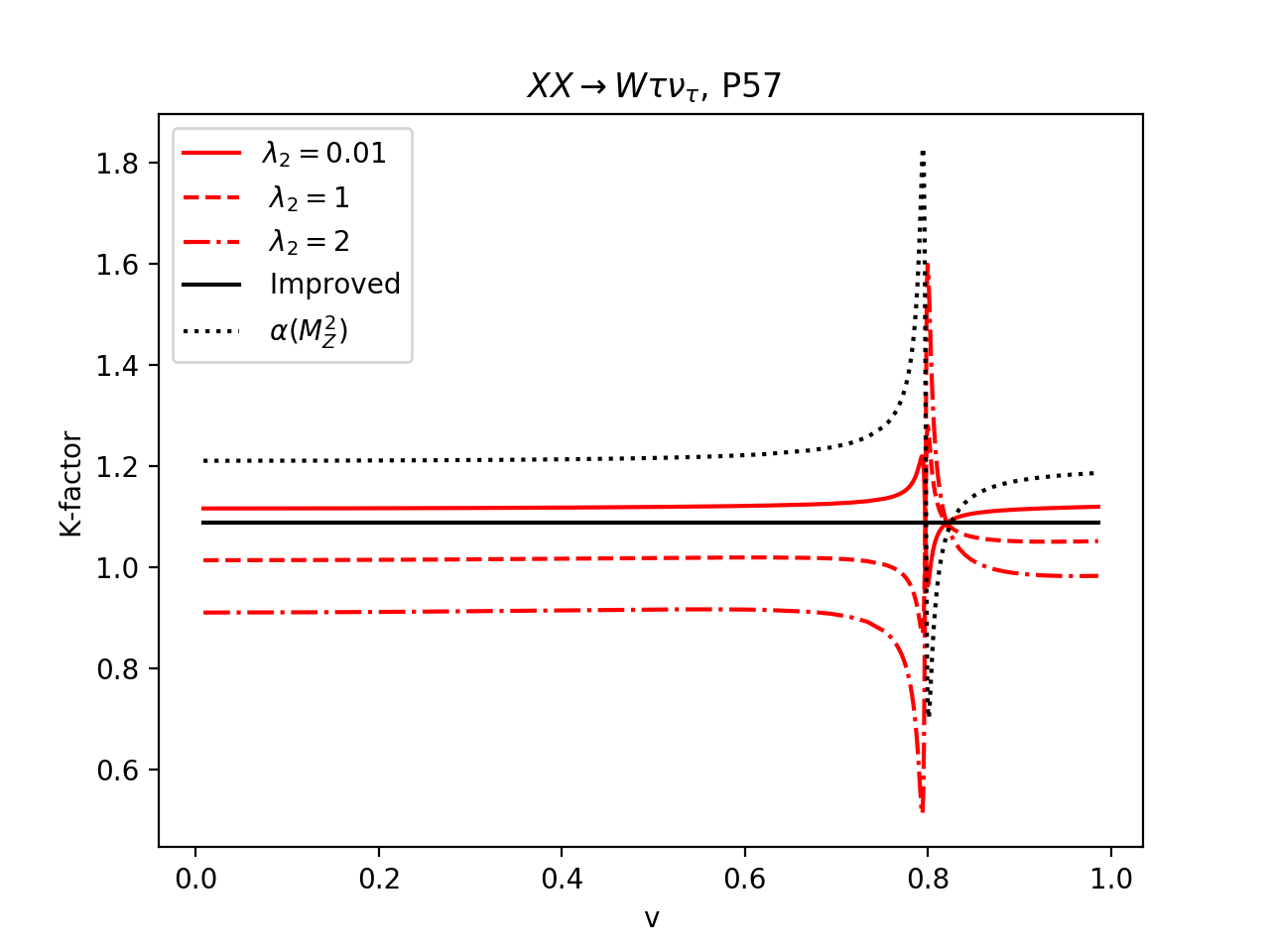}
\includegraphics[width=0.45\textwidth]{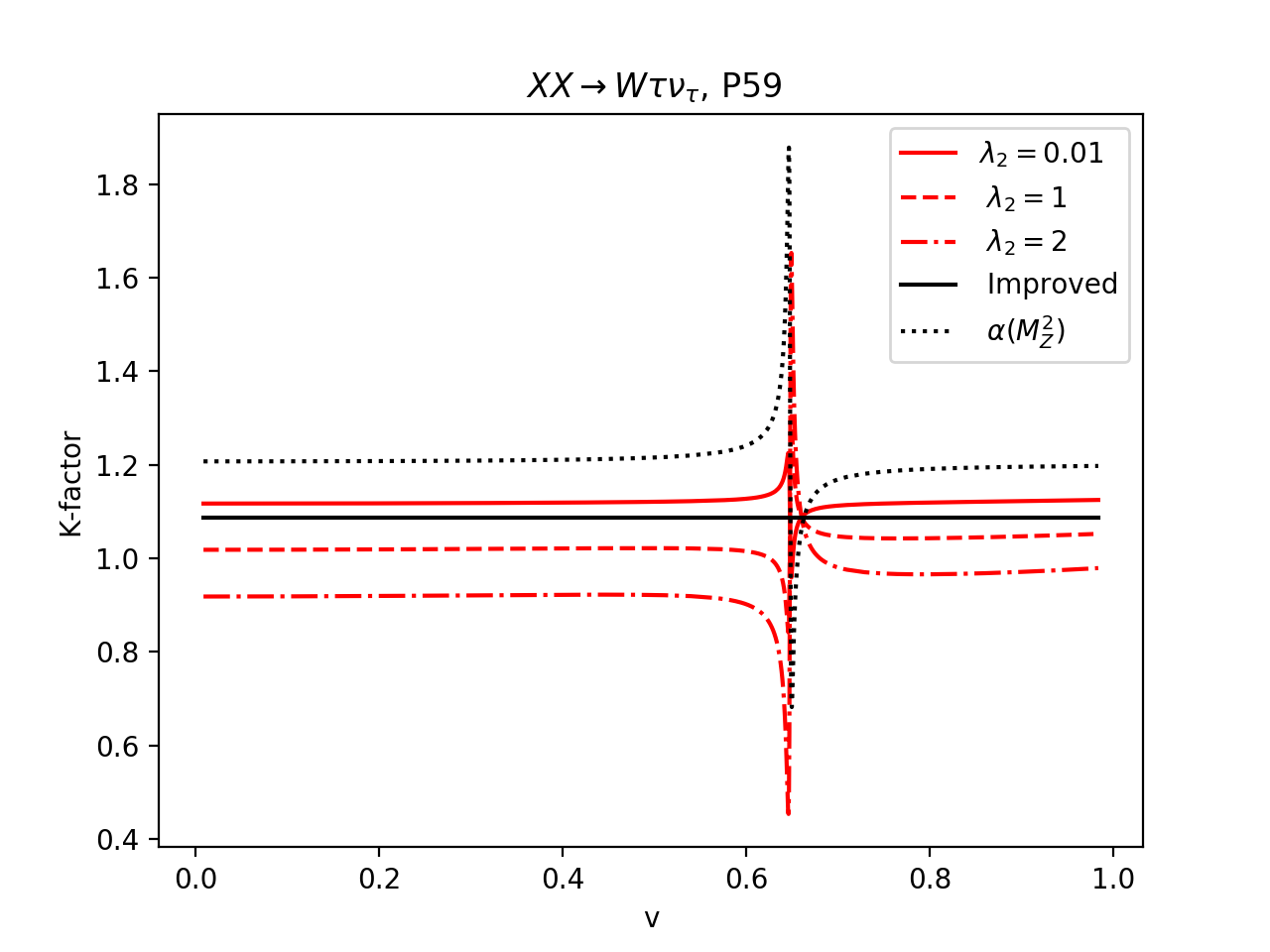}
\includegraphics[width=0.45\textwidth]{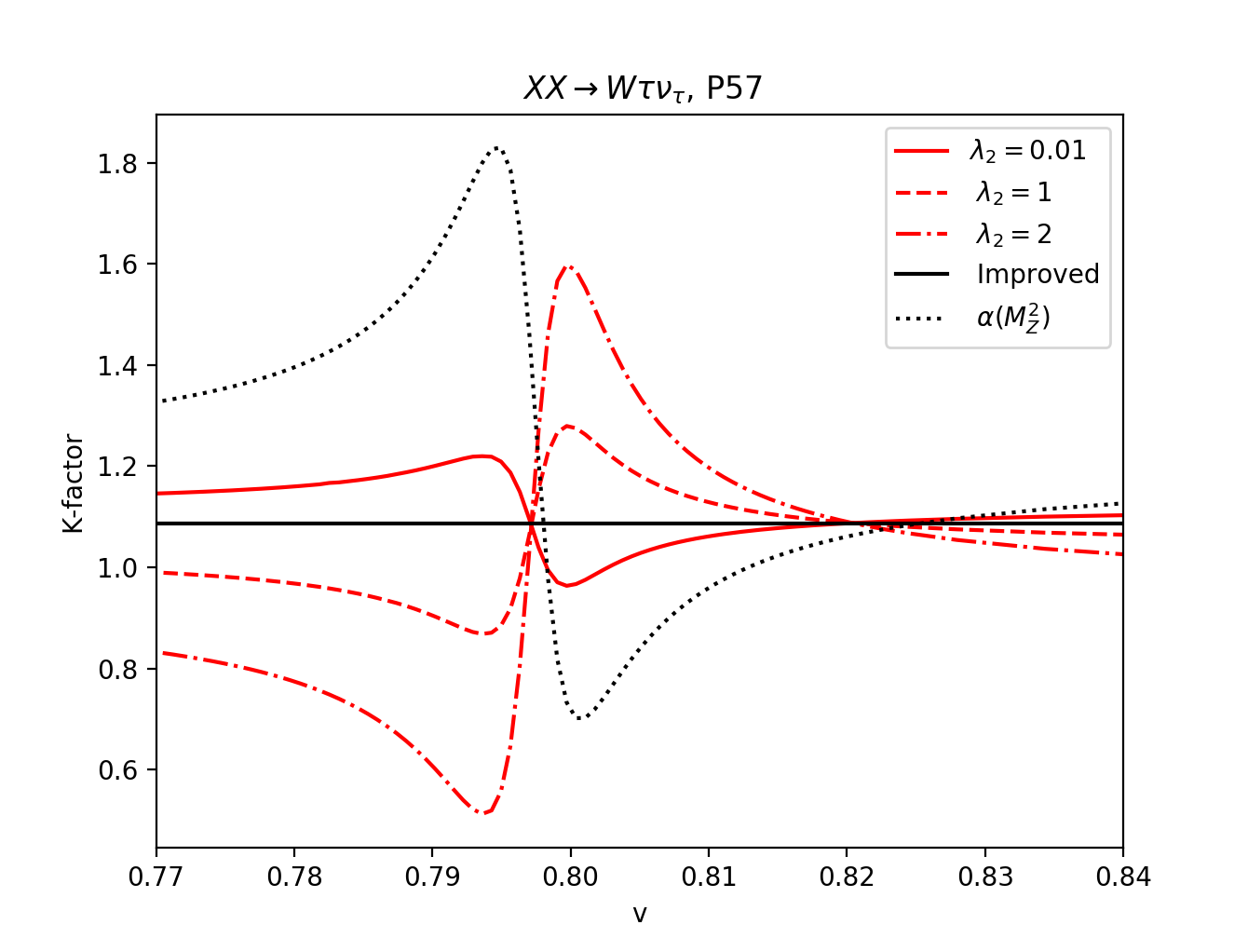}
\includegraphics[width=0.45\textwidth]{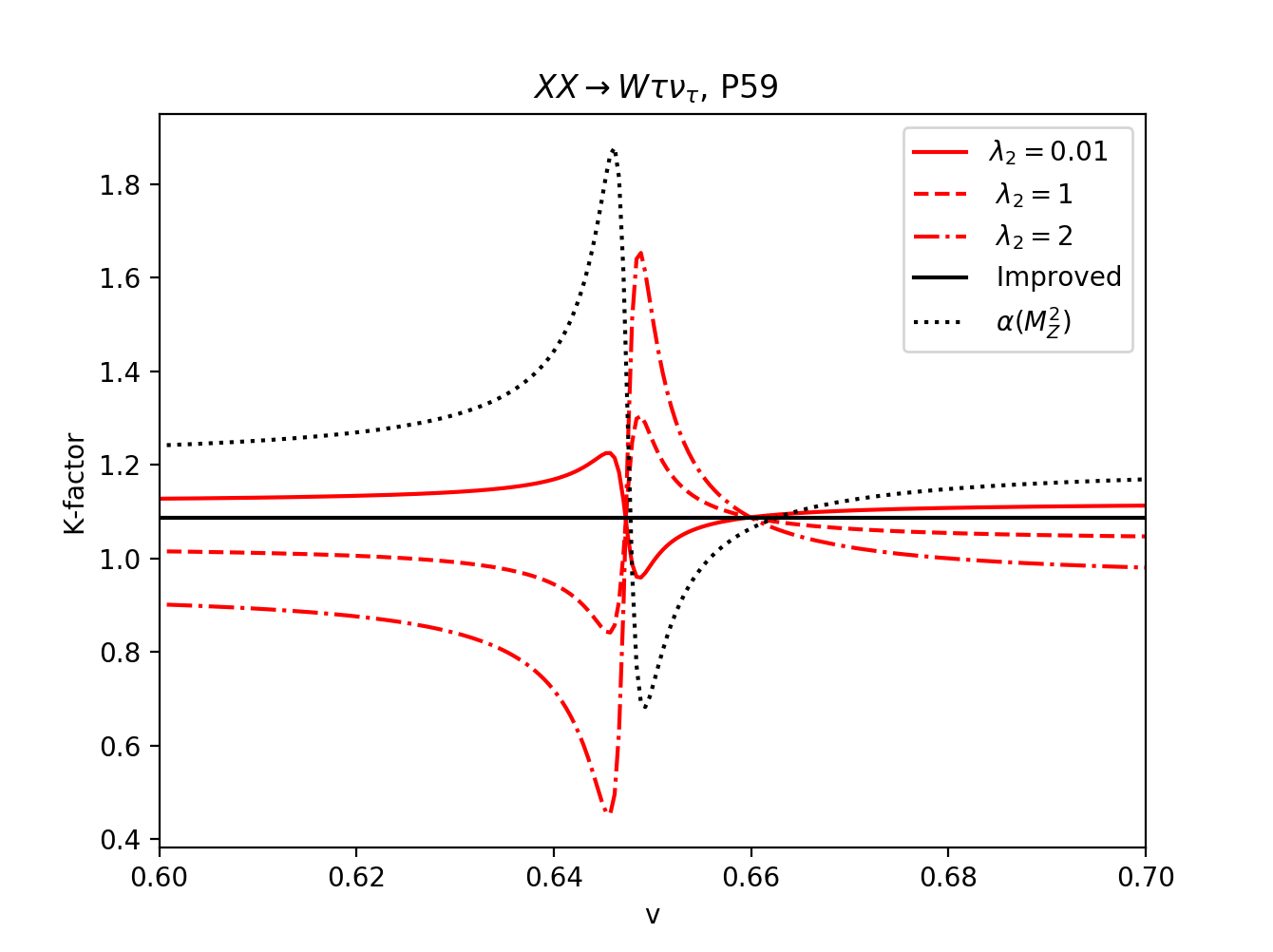}
\caption{\it  \label{fig:xxtoWlnlloopresults} As in Figure~\ref{fig:xxtobbloopresults} but for the $ X X \to W l \nu_l$ channel.}
\end{center}
\end{figure}

First of all, we confirm that the {\it relative} correction to the full cross-section over the whole range of $v$, is within $0.04\%$, independent of the fermion flavour. The weight of each channel equals the branching ratio of the $W$ into the corresponding fermion. We therefore only show our results for the lepton channel. These results are shown in Figure~\ref{fig:xxtoWlnlloopresults}. Were we to look at the total correction, we would see that the corrected cross-sections follow the tree-level cross-section, the very peaked resonance structure does not show the details of the corrections. The details of the corrections are most clearly revealed in the normalised (with respect to the tree-level result) corrections. Again, as in the case of $b \bar b$, apart from the location of the Higgs peak in terms of the relative velocity, $v \sim 0.82$ for P57, and $v \sim 0.66$ for P59, the characteristics for the 2 benchmark points are quite similar. At the Higgs peak the correction is very small. In fact, at this point, the improved tree-level agrees perfectly with the full one-loop correction, as expected. As explained previously, by construction, at $s=M_h^2$, the fact that $\Gamma(h \to XX)$ is used as an OS input, the $\l_2$ dependence of the correction is absorbed in the renormalisation and therefore the $\l_2$ dependence does not show up at the peak. Moreover, at the peak, our implementation of the width makes any non-resonant contribution vanish. Outside the peak, the $\l_2$ dependence is important and is magnified just around the peak in the interference region. For $\l_2=0.01$, the variation with $v$ is modest. This variation amplifies as $\l_2$ is increased. We also note that the sign of the interference changes with increasing positive $\l_2$ for both benchmark points. Observe also that even far away from the resonance, $v\sim 0$, the $\l_2$ dependence is quite different from the one we found in the $b \bar b$ channel. For $b \bar b$, $\l_2=0.01$ amounts to a reduction of the tree-level cross-section by about $10\%$ while here the same value of $\l_2$ gives an enhancement of about $20\%$ for small $v$. Also, the corrections for P59 are slightly smaller than for P57, again contrary to the $b \bar b$ channel. Another important difference from the $b \bar b$ channel is that the effective $\alpha(M_Z^2)$ does a good job only very close to the peak. In the region between the dip (anti-resonance) and the peak, where the interference between the Higgs channel and the gauge continuum contribution is important, the $\alpha(M_Z^2)$ is not a good approximation. It affects the Higgs contribution and the $WW^\star$ contribution differently. Moreover, for small $v$ the "$\alpha(M_Z^2)$" cross-section gives corrections which are about a factor 2 larger than the full loop correction for $\l_2=0.01$. The discrepancy worsens as $\l_2$ increases, confirming that $\l_2 \sim 0$ is what best approaches the "$\alpha(M_Z^2)$" approximation.  It is clear that the approximations do not reproduce the full one-loop correction (which we can equate with the result of $\l_2=0.01$) nor do they catch the $\l_2$ dependence. Will these important features have an important impact on the relic density, considering that the cross-sections are dominated by the peak contribution which in our implementation, these features are by construction set to zero? They may also be diluted because of the weight of this channel (compared to the $b \bar b$ channel) to the total effective annihilation cross-section and hence to the relic density. 


\section{Application to the relic density}
\label{sec:applicationtorelic}
One of the steps in the calculation of the relic density, given the velocity dependent annihilation cross-sections, requires integrating the convolution of those cross-sections with the velocity distribution. With very narrow resonance structures, the integrator must carefully pick up the peaked region. For the case of a resonance with a very nearby anti-resonance like the case of the $W f \bar f^\prime$ final state, the integrator must pick up the entirety of this rich structure. This is the reason we generate a very large set of points around the resonance, carefully including the peak and the dip as well as taking into account the contributions away from the resonance. This is important because our renormalisation condition together with a consistent implementation of the Higgs width show that the peak region is practically not corrected while the (relative) corrections get larger as soon as we move away from the exact resonance point. The integration over the velocity, for some values of $x_F=M_X/T_F$ ($T_F$ is the freeze-out temperature) is performed through two dedicated integrators, one calling {\tt Python} libraries and the other with {\tt Mathematica}. Moreover, analytical formulae when the test (cross-section) function is of the form $1/((s-M_h^2)^2+(\Gamma_h M_h)^2), (s-M_h^2)/((s-M_h^2)^2+(\Gamma_h M_h)^2), 1$~\footnote{These functions correspond, respectively, to the resonance, the interference term and the continuum.} with the convolution over the velocity distribution $v^2 e^{(-x_F v^2)}$ exist when the integration is carried over the range $v={0, \infty}$. The analytical formulae are then compared with the numerical evaluation of the dedicated integrators. Only then do we first feed {\tt micrOMEGAs} these test functions to conduct sanity checks and then feed in the exact one-loop cross-sections resulting from our code, {\tt SloopS}. Of course {\tt micrOMEGAs} solves also the Boltzmann equation and returns the effective $x_F$ which we use for the integrators to conduct our checks. 
\begin{table}[hb!]\begin{center}
\begin{tabular}{c|c|c|c|c|c|c|}
\cline{2-7}
 & Tree & $\l_2$ = 0.01 & $\l_2$ = 1 & $\l_2$ = 2 & Improved &  $\alpha(M_Z^2)$ \\ \hline
\multicolumn{1}{|c||}{{\bf P57}}& 0.113 & 0.110 (-2.42\%) & 0.111 (-2.21\%) & 0.111 (-2.00\%) & 0.110 (-2.36\%)) & 0.118 (4.43\%) \\ \hline
\multicolumn{1}{|c||}{{\bf P59}}& 0.108 & 0.105 (-2.73\%) & 0.105 (-2.37\%) & 0.106 (-2.00\%) & 0.105 (-2.62\%)  & 0.113 (4.63\%)\\ 
\hline \hline
\end{tabular}
\end{center}
\caption{\it Correction to the relic density for the Higgs resonance Points P57 and P59.}
\label{tab:rel-cont-P57-59}
\end{table}

The corrections to the relic density for P57 and P59 are shown in Table~\ref{tab:rel-cont-P57-59}. The corrections are very small, of the order $-(2-3)\%$. These small corrections are a consequence of the fact that the cross-sections are overwhelmingly dominated by the cross-section at the resonance, $s=M_h^2$, for which our implementation of the loop corrections suppresses all non-Higgs mediated contributions and gives a correction to the $b\bar b$ cross-section of order $1\%$ (only slightly more for $WW$). The corrections away from the peak, though large in relative terms, are small in absolute terms. Their effect is {\it detectable} and in principle (at the per-mille level) the $\l_2$ dependence is traceable. Note that the use of $\alpha(M_Z^2)$ gives larger (but small positive) corrections. This is due to the fact that the $b \bar b$ channel is independent of the choice of $\alpha$ as explained in the previous section whereas while its effect is large in $WW$, its effects are much reduced because of the much smaller weight of this channel in the total effective cross-section. 


\section{Conclusions}
\label{sec:conclusions}
We presented a very thorough procedure for the implementation of the width at tree-level and at one-loop that avoids the problem of double counting and eventually of the breakdown of unitarity especially in a situation like the $WW^\star$ final state where interference between the resonant contribution and other amplitudes is important. It is crucial to stress that the fact that the one-loop corrections are very small at the resonance is due to, on the one hand, the implementation of the fudge factor to maintain gauge invariance (for $WW^\star$ this maintains the correct $\l_L$ dependence) but sets to zero any non resonant amplitude, and on the other hand,  the use of the OS renormalisation scheme which entails that the correction to the vertex $hXX$ is zero at the resonance. Away from the resonance, these corrections, in relative terms, are not small. We hope that our analysis of the resonance and the procedure we detailed proves helpful for many applications, not necessarily related to dark matter. It rests that for our applications to the Higgs resonance, an extremely narrow resonance, the very small correction at the resonance within our scheme and procedure means a very small correction of the velocity integrated cross-sections and hence the relic density. Because the integration over the resonance needs to pick up all the very subtle changes in the cross-sections, we performed sanity checks that ensure that the numerical integration is performed correctly. 

\acknowledgments
We thank Alexander Pukhov for several helpful discussions. HS is supported by the National Natural Science Foundation of China (Grant No.12075043, No.11675033). He thanks the CPTGA and LAPTh for support during his visit to France in 2019 when this work was initiated. SB is grateful for the support received from IPPP, Durham, UK, where most of this work was performed. SB also acknowledges the support received from LAPTh, during his visit, when this work was started. NC is financially supported by IISc (Indian Institute of Science, Bangalore, India) through the C.V.~Raman postdoctoral fellowship. NC also thanks the support received from DST, India, under grant number IFA19-PH237 (INSPIRE Faculty Award).

\bibliography{../../NLO_IDM570}

\bibliographystyle{JHEP}

\end{document}